\documentclass[12pt]{article}
\usepackage{axodraw,epsf}
\usepackage{amssymb,color}
\usepackage{cite}
\usepackage{graphicx}

\newcommand{\be}{\begin{equation}}
\newcommand{\ee}{\end{equation}}
\newcommand{\bea}{\begin{eqnarray}}
\newcommand{\eea}{\end{eqnarray}}
\newcommand{\fd}{{F_{2}^{D(3)}}}

\newcommand{\ftwopom}{{F_{2}^{\pom}}}
\newcommand{\pe}{{p_{e}}}
\newcommand{\pep}{{p_{e}^{\prime}}}
\newcommand{\pa}{{p_{A}}}
\newcommand{\pb}{{p_{B}}}
\newcommand{\gammast}{{\gamma^{*}}}
\newcommand{\betast}{{\beta^{*}}}
\newcommand{\thetaone}{{\theta_{1}}}
\newcommand{\thetatwo}{{\theta_{2}}}
\newcommand{\thetatwop}{{\theta_{2}^{\prime}}}
\newcommand{\degg}{{^{\circ}}}
\newcommand{\cms}{{^{\rm cms}}}
\newcommand{\lab}{{^{\rm lab}}}
\newcommand{\ha}{{\frac{1}{2}}}
\newcommand{\pom}{{I\!\!P}}
\newcommand{\xpom}{{x_{I\!\!P}}}
\newcommand{\intercept}{{\alpha_{I\!\!P}(0)}}
\newcommand{\mxsq}{{M_{X}^{2}}}
\newcommand{\qsq}{{Q^{2}}}
\newcommand{\ptsq}{{p_{\bot}^{2}}}
\newcommand{\ptsqmin}{{p_{\bot\,{\rm min}}^{2}}}
\newcommand{\kmin}{{k^2_{\rm min}}}
\newcommand{\etamax}{{\eta_{\rm max}}}
\newcommand{\etamaxexp}{{\eta_{\rm max}^{\rm expt}}}
\newcommand{\etamaxth}{{\eta_{\rm max}^{\rm th}}}

\begin{document}
\title{\hfill{\small CERN-TH/98-396} \\
\vspace*{-0.3cm}
\hfill{\small OUTP-98-91P} \\
\bigskip
Probing the Structure of the Pomeron}
\author{{\bf John Ellis } \\
Theoretical Physics Division, \\
CERN,\\
1211 Geneva 23,\\
Switzerland\\
{\bf Graham G. Ross} and {\bf Jenny Williams}\\
Dept. of Physics,\\
Theoretical Physics,\\
1 Keble Rd., \\
Oxford OX1 3NP}
\date{\today}
\maketitle

\begin{abstract} 
We suggest that the pseudo-rapidity cut dependence of diffractive
deep-inelastic scattering events at HERA may provide a sensitive test of
models of diffraction. A comparison with the experimental cross
section shows that the Donnachie-Landshoff model and a simple two-gluon
exchange model of the pomeron model are disfavoured. However a
model with a direct coupling of the pomeron to quarks is viable for a
harder quark--pomeron form factor, as is a model based on the 
leading-twist operator contribution. We also consider a direct-coupling
scalar pomeron model. We comment on the implications of
these results for the
determination of the partonic structure of the pomeron.
\end{abstract}

\pagestyle{empty}

\setcounter{page}{1} \pagestyle{plain}

\section{Introduction}

\label{sect:intro}

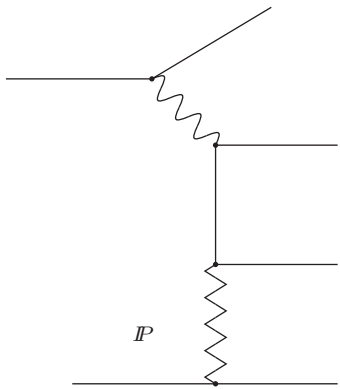
\begin{figure}[htp]
\begin{center} \begin{picture}(180,160)(-20,0)
\SetScale{1}
\Line(5,128)(60,128)
\Line(60,128)(105,155)
\Line(84,103)(130,103)
\Line(84,58)(130,58)
\Line(84,103)(84,58)
\Line(30,13)(130,13)
\Photon(60,128)(84,103){4}{3.5}
\Text(57,32)[]{\scriptsize $\pom$}
\ZigZag(84,58)(84,13){-4}{4.5}
\Vertex(60,128){1}
\Vertex(84,103){1}
\Vertex(84,58){1}
\Vertex(84,13){1}
\end{picture}
\caption[Diffractive DIS via pomeron exchange.]{{\it Diffractive 
DIS via pomeron exchange.}\label{fig:dijet}}
\end{center}
\end{figure}

\begin{figure}[htp]
\begin{center} \begin{picture}(260,100)(65,0)
\SetScale{0.5}
\Text(3,-5)[lt]{(a)}
\Line(5,128)(60,128)
\Line(60,128)(105,155)
\Line(84,103)(130,103)
\Line(84,58)(130,58)
\Line(84,103)(84,58)
\Line(30,13)(130,13)
\Photon(60,128)(84,103){4}{3.5}
\SetColor{Blue}
\ZigZag(84,58)(84,13){-4}{4.5}
\SetColor{Red}
\Gluon(99,103)(130,80){-4}{3.5} 
\SetColor{Black}
\Vertex(60,128){1}
\Vertex(84,103){1}
\Vertex(84,58){1}
\Vertex(84,13){1}
\Vertex(99,103){1} 
\Line(175,128)(230,128)
\Line(230,128)(275,155)
\Line(254,103)(300,103)
\Line(254,58)(300,58)
\Line(254,103)(254,58)
\Line(200,13)(300,13)
\Photon(230,128)(254,103){4}{3.5}
\SetColor{Blue}
\ZigZag(254,58)(254,13){-4}{4.5}
\SetColor{Red}
\Gluon(254,80)(300,80){4}{4.5}
\SetColor{Black}
\Vertex(230,128){1}
\Vertex(254,103){1}
\Vertex(254,58){1}
\Vertex(254,13){1}
\Vertex(254,80){1}
\Line(345,128)(400,128)
\Line(400,128)(445,155)
\Line(424,103)(470,103)
\Line(424,58)(470,58)
\Line(424,103)(424,58)
\Line(370,13)(470,13)
\Photon(400,128)(424,103){4}{3.5}
\SetColor{Blue}
\ZigZag(424,58)(424,13){-4}{4.5}
\SetColor{Red}
\Gluon(439,58)(470,80){4}{3.5} 
\SetColor{Black}
\Vertex(400,128){1}
\Vertex(424,103){1}
\Vertex(424,58){1}
\Vertex(424,13){1}
\Vertex(439,58){1} 
\Text(285,-5)[lt]{(b)}
\Line(555,128)(610,128)
\Line(610,128)(655,155)
\Line(634,103)(680,103)
\Line(634,73)(680,73)
\Line(634,103)(634,73)
\Line(580,13)(680,13)
\Photon(610,128)(634,103){4}{3.5}
\SetColor{Blue}
\ZigZag(634,43)(634,13){-4}{3.5}
\SetColor{Red}
\Gluon(634,43)(634,73){4}{3.5}
\Gluon(634,43)(680,43){4}{4.5}
\SetColor{Black}
\Vertex(610,128){1}
\Vertex(634,103){1}
\Vertex(634,73){1}
\Vertex(634,43){1}
\Vertex(634,13){1}
\end{picture}
\vspace*{.7cm}
\caption[Higher order QCD corrections to diffractive scattering via
pomeron exchange]{{\it Higher-order QCD corrections to diffractive
scattering via pomeron exchange, due to (a) gluon bremsstrahlung and (b)
boson-gluon fusion diagrams.}\label{fig:3jets}}
\end{center}
\end{figure}
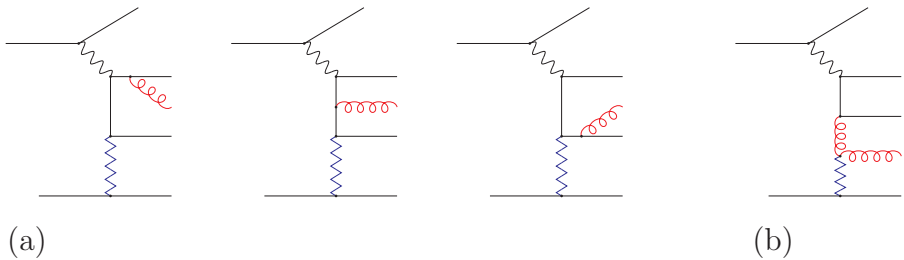

Pomeron exchange is familiar as a description of total
hadron-hadron cross sections that rise slowly with increasing
energy\,\cite{Donnachie:1984hf:Donnachie:1984xq:Donnachie:1986iz:Donnachie:1992ny}.
It should also play a major role in the diffractive events that have
been observed in electron-proton deep-inelastic scattering (DIS) at HERA,
where the diffractive system may be isolated by requiring that it be 
separated from the proton direction by a large gap in
pseudo-rapidity\,\cite{Ahmed:1994nw:Ahmed:1995ui:Derrick:1993xh,Derrick:1994ze}.
Large pseudo-rapidity gaps imply the exchange of a colourless state
between the proton and the virtual photon, and the leading
contribution to these events is often interpreted as due to pomeron
exchange\,\cite{Donnachie:1984hf:Donnachie:1984xq:Donnachie:1986iz:Donnachie:1992ny}.
These pseudo-rapidity gap events can be viewed as the proton emitting
a pomeron which then undergoes deep-inelastic scattering. Ingelman and
Schlein\,\cite{Ingelman:1985ns} suggested treating the pomeron as an
hadronic particle. In this picture, one may consider diffractive DIS
to be a probe of the partonic structure of the pomeron. This is a
{\em resolved-coupling} scheme, as the high-energy photon sees the
constituent partons in the pomeron. That is, the quarks and gluons to
which the pomeron couples are considered to be constituents of the
pomeron, and hence are necessarily close to mass shell.  The basic
diagram which illustrates this scheme is shown
in~Fig.\,\ref{fig:dijet}, where one may imagine that the photon is probing
the quark structure of the pomeron. Higher-order processes include
gluon bremsstrahlung, as shown in Fig.\,\ref{fig:3jets}a, and
boson-gluon
fusion, as shown in Fig.\,\ref{fig:3jets}b, in which the photon
interacts with
the gluonic structure of the pomeron. The resolved-coupling
interpretation does not include direct couplings between the pomeron and
off-shell partons.

In this hadronic picture of the pomeron, one can factorize the
diffractive scattering cross section into a sum over the product of
the probability to find a parton with momentum fraction~$\beta$ in the
pomeron and the cross section for parton-photon hard scattering. In
analogy with standard DIS, one can define a diffractive structure
function for these processes in terms of the diffractive cross
section,

\be \label{eq:sfn}
\frac{{\rm d}^3 \sigma^{\rm diff}}{{\rm d}\beta\,{\rm
d}Q^2\,dx_{I\!\!P}}=
\frac{2\pi\alpha^2}{\beta Q^4} [(1+(1-y)^2)F_2^{D(3)}(\beta ,Q^2,x_{I\!\!P}) -
y^2F_{L}(\beta ,Q^2,x_{I\!\!P})],
\ee

\noindent where for~$y\lesssim 0.4$ the longitudinal term can be neglected\,
\cite{Ahmed:1995ns}. According to this picture
for leading pomeron exchange, the diffractive structure function can be
written in terms of a {\em pomeron structure function},~$F_{2}^{I\!\!P}(%
\beta,Q^2)$,

\be \label{eq:pomsfn}
F_{2}^{D(3)}(\beta
,Q^2,x_{I\!\!P})=f^{I\!\!P}(x_{I\!\!P})\,F_{2}^{I\!\!P}(\beta,Q^2).
\ee

\noindent Here~$f^{I\!\!P}$ describes the pomeron flux in the
proton,~$\beta$ plays the role of the Bjorken $x$ variable in DIS,
and~$F^{I\!\!P}$ is interpreted as the structure function of the
pomeron. Experimental results from HERA are consistent with this
factorization for a wide range of
parameters\,\cite{Breitweg:1998aa}. Factorization breaking can be
interpreted as being due to sub-leading
Reggeon~($\rho$,~$\omega$\,\ldots) exchanges\,\cite{Adloff:1997sc}.

One consequence of a strong pseudo-rapidity cut
selection\,\cite{Ellis:1996cg} is that it tends to force the parton
far off-shell, introducing a further possible source of factorization
breaking into the diffractive structure function. This is because the
experimentally extracted diffractive structure function is obtained by
integrating the differential cross section over the kinematically
accessible phase space. As shown in Appendix\,A.1, the lower limit
$\theta_{\rm min}^{\rm cms}$ of the angular integral is a sensitive
function of the kinematic parameters $\qsq$, $\beta$ and
$\xpom$. Therefore, as we discuss in Sect.\,\ref{sect:tests}, a
further, unfactorizable, $\xpom$-dependence is introduced into the
diffractive structure function through the lower limit of the
phase-space integral. As a result, one would not expect a diffractive
structure function selected by making a large pseudo-rapidity cut,
requiring large virtuality, to be factorizable into a pomeron flux
factor and an $\xpom$-independent pomeron ``structure function'', even
though the full structure function (integrated over the whole phase
space) may be factorizable.

As discussed in~\cite{Ellis:1996cg}, events with a strong
pseudo-rapidity cut offer a way of discriminating between models with
direct and resolved couplings. Here we sharpen the previous analysis
of the kinematical constraints implied by pseudo-rapidity gap cuts,
and discuss how they may offer a sensitive discriminator between
various models proposed to describe deep-inelastic diffractive
scattering. We present a comparison with experiment of the
pseudo-rapidity dependence for five models. The first is the
Donnachie-Landshoff (D-L)
model\,\cite{Donnachie:1987pu:Donnachie:1987xh}, which employs a
vector-like direct coupling of the Pomeron to quarks, assuming a soft
form factor. The second is a modification of this scheme suggested by
two of the authors (E-R)\,\cite{Ellis:1996cg}, which invokes a harder
form factor. The third model assumes a scalar direct coupling to
quarks\,\cite{Vermaseren:1996iy}. The fourth model uses two-gluon
exchange to model the Pomeron\,\cite{Diehl:1995wz}, and the final
model assumes that the pseudo-rapidity gap events can be described by
the operator-product expansion (OPE), with dominance by the
leading-twist operator. Finally we discuss the implications of our
analysis for the prospects of probing the parton structure of the
Pomeron.

In Sect.\,\ref{sect:kinematics} we describe the kinematics of
diffractive deep-inelastic electron-proton scattering. Following this,
in Sect.\,\ref{sect:virtuality} we discuss the parton virtuality
constraints implied by strong pseudo-rapidity cuts, the derivations of
which are given in Appendices\,A.1 and~A.2. In Sect.\,\ref{sect:ope} we
discuss our selection of the various models that have been proposed to
describe
diffractive deep-inelastic scattering. In Sect.\,\ref{sect:tests} we
compare the predictions of these models for the pseudo-rapidity gap
dependence with experiment. Finally, we look at the consequences of
large virtuality constraints for previous analyses and suggest in
Sect.~6 further
experimental investigations to test the ideas presented here. A
summary and conclusions are presented in Sect.~7.

\section{Kinematics of Diffractive Deep-Inelastic Scattering}

\label{sect:kinematics}

In the HERA electron-proton experiments, 820\,GeV protons collide with
27.5\,GeV electrons or positrons. This corresponds to a centre-of-mass
(CMS) energy $\sqrt{s}\sim300$\,GeV, and allows access to a wider range
of~$Q^2$ and~Bjorken $x$ than has previously been possible.
In the HERA lab frame, the positive~$z$ axis is defined to be in the
forward proton direction and the origin is at the interaction vertex.
We consider diffractive deep-inelastic $e - P$ scattering,

\be
e(p_{e}) + P(P) \rightarrow e(p_{e}^{\prime}) + P(P^{\prime}) + X(X),
\ee

\noindent where the momenta of the particles are shown in
brackets. The hadronic system~$X$ is assumed to be
separated from the forward proton
direction by a large pseudo-rapidity gap.
We assume that the proton (or low-mass
excited state) escapes undetected down the beam-pipe or is detected
far downstream\,\cite {Breitweg:1998aa}. The contribution from proton
dissociation is limited by acceptance cuts: if~$M_{Y}$ is the mass of
the proton remnant, then for H1~$M_{Y}\!\lesssim\! 1.6\,{\rm GeV}$,
and for ZEUS~$M_{Y}\!\lesssim\! 4\,{\rm GeV}$. One may consider that the
interaction proceeds by virtual photon--pomeron deep-inelastic
scattering,

\be
\gamma^{*}(q) + I\!\!P(P_{I\!\!P}) \rightarrow X(X),
\ee

\noindent where~$P_{I\!\!P}=P-P^{\prime}$.

We use the usual kinematic variables of deep-inelastic scattering:

\be
Q^2=-q^2,~~~x=\frac{Q^2}{2P\cdot q},~~~{\rm and}~~y=\frac{Q^2}{x\,s},
\ee

\noindent where~$Q^2$ is the negative 4-momentum squared of the virtual
photon and~$x$ is the Bjorken scaling variable. We also define~$W^2$, the
mass squared of the total hadronic system ($X\,+\,$outgoing proton), by

\be
W^2=(P+q)^2.
\ee

\noindent Additionally, for diffractive scattering we define

\be
t_{I\!\!P}=(P-P^{\prime})^2,~~~x_{I\!\!P}=
\frac{(P-P^{\prime})\cdot q}{P\cdot q} \approx
\frac{Q^2+M_{X}^2}{Q^2+W^2},~~~{\rm
and}~~\beta=\frac{Q^2}{2(P-P^{\prime})\cdot q} \approx
\frac{Q^2}{Q^2+M_{X}^2}, \ee

\noindent where~$t_{I\!\!P}$ is the momentum transfer at the proton vertex
and is constrained by experimental cuts to be small ($%
|t_{I\!\!P}|\lesssim 1$\,GeV$^2$),~$x_{I\!\!P}$ is the fraction of
longitudinal momentum of the proton carried by the pomeron,\footnote{%
To a good approximation for very small~$t_{I\!\!P}$, the pomeron is emitted
in the forward proton direction.} and~$x=\beta x_{I\!\!P}$. The mass squared
of the diffractive system~$X$ is~$M_{X}^2$, and the proton mass is neglected
in this analysis. In the lowest-order diagram shown in
Fig.\,\ref{fig:dijet}, $\beta$%
~is interpreted as the fraction of the pomeron momentum carried by the
struck quark, whereas in the three-jet diagrams of
Fig.\,\ref{fig:3jets},~$\beta$ is
the fraction of pomeron momentum in the exchanged parton which couples to
the pomeron.

The {\em pseudo-rapidity}~$\eta$ of an outgoing particle
is defined in the laboratory frame in terms of its polar angle with
respect to the proton direction:

\be
\eta=-\ln\tan\left(\frac{\theta_{\rm lab}}{2}\right).
\ee

\noindent
In the Ingelman-Schlein picture~\cite{Ingelman:1985ns}, diffractive DIS
corresponds to probing the
partonic structure of the pomeron. For example, the leading
contribution to diffractive scattering is the dijet diagram
of~Fig.\,\ref{fig:dijet}, in which the high-energy photon sees the quark
content of the pomeron. Higher-order processes include the gluon
bremsstrahlung of~Fig.\,\ref{fig:3jets}a and the boson-gluon
fusion of~Fig.\,\ref{fig:3jets}b, 
in which the photon interacts with the gluonic structure of
the pomeron.

For the process of~Fig.\,\ref{fig:dijet}, we introduce a further invariant,
the four-momentum squared of the struck quark,~$k^2$. In the~$%
\gamma^{*}-I\!\!P$ CMS system, the virtuality of this quark
can be expressed in terms of other invariants, and the polar angle with
respect to the~$\gamma^{*}\,I\!\!P$ axis, by

\be
k^2= -\frac{Q^2+M_{X}^2}{2}(1-\cos\theta_{\rm cms}).  \label{eq:ksqdefn}
\ee

\noindent A similar expression can be formed for interactions, such as
those
of~Fig.\,\ref{fig:3jets}, where more than two final-state partons are
produced.

\section{Virtuality Constraints from Experimental Cuts}
\label{sect:virtuality}

In a typical measurement with a pseudo-rapidity cut,
diffractive events are selected by requiring there to be no activity
observed above a low-energy threshold (400\,MeV) in a large
pseudo-rapidity interval
about the forward proton direction. Thus only events with pseudo-rapidity
less than some cut,~$\eta_{\rm max}$, are accepted.
On the other hand, for an hadronic interpretation of the pomeron, we
would require the quarks
coupling to the pomeron to be near mass
shell~($|k^2|\lesssim\Lambda^2_{\rm QCD}$).  
It has been observed that the strong
pseudo-rapidity cuts imposed in early H1 and ZEUS
analyses\,\cite{Ahmed:1994nw:Ahmed:1995ui:Derrick:1993xh,Derrick:1994ze}
restricted the phase space available for diffractive scattering, and
selected only events in which the struck quark of~Fig.\,\ref{fig:dijet}
was forced to be far off mass shell for a wide region of parameter
space\,\cite{Ellis:1996cg}. This means that the selection cuts rejected
all events corresponding to the process of~Fig.\,\ref{fig:dijet} for a
wide range of parameters. 

To demonstrate this, note that a cut in
pseudo-rapidity corresponds to a lower bound on~$\theta _{\rm cms}$,
which translates to a lower bound,~$k_{\rm min}^{2}$, on the struck
quark virtuality. As was shown previously\,\cite{Ellis:1996cg}, for
strong pseudo-rapidity cuts $\eta _{\rm
max}$, and for a wide range of~$\beta $,~$Q^{2}$ and~$x_{I\!\!P}$, the
struck quark in the diagram of~Fig.\,\ref{fig:dijet} is forced to be
far off~shell,~i.e.,~$-k_{\rm min}^{2}>1\,{\rm GeV}^{2}$.%
\footnote{We note that the experimental cuts are made at the hadron
level, and that our calculations are at parton level. The
corresponding estimate of~$\eta _{\rm max}$ at the parton level uses a
conservative estimate of the hadronization
radius\,\cite{Derrick:1994ze}, namely, it assumes that hadronization
radius spans approximately half a unit in pseudo-rapidity. We have
tested the robustness of our calculations by also considering the extreme
cases where
the hadronizing parton is at either edge of the resulting jet,
finding that a large lower bound, $-k_{\rm min}^{2}\gtrsim 1\,{\rm
GeV}^{2}$, on the exchanged quark virtuality remains.} The result
is a relation between~$\eta _{\rm max}$ and~$k_{\rm min}^{2}$ in terms
of the laboratory energies of the electron and proton and the kinematic
variables defined in the previous section. The details of this
calculation are given in~Appendix\,A.1. Apart from correcting a small
error, the main difference between the new and old
calculations~\cite{Ellis:1996cg} is due to
the elimination of a small-angle approximation made in the original
analysis.

The virtuality constraints following from the pseudo-rapidity cuts
used in early ZEUS analyses are given in
Table\,\ref{table:ervirtuality},
following~\cite{Ellis:1996cg}. Tabulated diffractive structure
function data for a pseudo-rapidity cut of $\etamax=1.8$ and much
stronger virtuality constraints appear in~\cite{Phillips:1995jpp}. The
kinematic parameters of these data
and the corresponding virtuality constraints, $\kmin$, are shown in
Table\,\ref{table:phillipsvirtuality}. This clearly demonstrates that,
in a resolved-coupling picture, events due to the process
of~Fig.\,\ref{fig:dijet} do not contribute to large pseudo-rapidity
gap diffractive DIS in a wide region of parameter space.

\begin{table}[htp]
\centering
\scriptsize{
\begin{tabular}{|c|l|c|r|l|c|c|}
\hline
&  &  &  &  &  &  \\ 
$Q^2$ & ~~~$\beta $ & $x_{I\!\!P}$ & 
$-k_{{\rm min}}^2\,{\rm GeV}^2$ & $-k_{{\rm min}}^2\,{\rm GeV}^2$ 
& 
$-k_{{\rm min\,old}}^2\,{\rm GeV}^2$ & $-k_{{\rm min\,old}}^2\,{\rm GeV}^2$ \\ 
${\rm GeV}^2$&  &  & $(\eta _{\rm max}=1.5)$ & $(\eta _{\rm max}=2.5)$
  & $(\eta _{\rm max}=1.5)$ & $(\eta _{\rm max}=2.5)$ \\ 
&  &  &  &  &  &  \\ \hline
&  &  &  &  &  &  \\ 
10 & 0.175 & ~.0032~ & 2.5 ~~~~~ & ~~~~~~0.39 & 3.1 & 0.4\\ 
&  & ~.0050~ & 5.1 ~~~~~ & ~~~~~~0.89 & 7.5 & 1.0\\ 
& 0.375 & ~.0020~ & 0.7 ~~~~~ & ~~~~~~0.11 &0.9 &0.12\\ 
&  & ~.0032~ & 1.5 ~~~~~ & ~~~~~~0.27 &2.3 &0.3\\ 
& 0.65 & ~.0013~ & 0.2 ~~~~~ & ~~~~~~0.03 &0.2 &0.03\\ 
&  & ~.0020~ & 0.3 ~~~~~ & ~~~~~~0.06 &0.5 &0.07\\ 
28 & 0.175 & ~.0050~ & 6.2 ~~~~~ & ~~~~~~0.95 &7.5 &1.0\\ 
&  & ~.0079~ & 13.3~~~~~ & ~~~~~~2.26 &18.7 &2.5\\ 
& 0.375 & ~.0020~ & 0.8 ~~~~~ & ~~~~~~0.12 &0.9 &0.1\\ 
&  & ~.0079~ & 7.4 ~~~~~ & ~~~~~~1.51 &14.2 &1.9\\ 
& 0.65 & ~.0020~ & 0.4 ~~~~~ & ~~~~~~0.06 &0.5 &0.07\\ 
&  & ~.0050~ & 1.7 ~~~~~ & ~~~~~~0.33 &3.2 &0.4\\ 
63 & 0.375 & ~.0050~ & 4.5 ~~~~~ & ~~~~~~0.71 &5.7 &0.8\\ 
&  & ~.0079~ & 9.4 ~~~~~ & ~~~~~~1.66 & 14.2&1.9\\ 
& 0.65 & ~.0032~ & 1.0 ~~~~~ & ~~~~~~0.16 &1.3 &0.2\\ 
&  & ~.0079~ & 4.1 ~~~~~ & ~~~~~~0.83 & 8.0&1.1\\ \hline
\end{tabular}}
\caption[Struck quark virtuality constraints for early ZEUS
data.]{{\it Constraints on the virtuality of the struck quark from the
pseudo-rapidity cuts used in early ZEUS
experiments\,\protect\cite{Derrick:1994ze}. In this Table,
$k_{\rm min\,old}^{2}$
corresponds to the results published in~\protect\cite{Ellis:1996cg},
and $k^2_{\rm min}$ are the revised
virtuality constraints with small corrections due principally to the
elimination of the small-angle approximation used
in~\protect\cite{Ellis:1996cg}.}}
\label{table:ervirtuality}
\end{table}

\begin{table}[htp] 
\centering		 
\scriptsize{
\begin{tabular}{|l|l|c|r|l|l|c|}
\hline
&  &  &  & & \\ 
$Q^2\,{\rm GeV}^2$ & ~~~~$\beta~\,~~ $ & ~~~~~$x_{I\!\!P}$~~~~~ &
$-k_{{\rm min}}^2\,{\rm GeV}^2$ & $-k_{{\rm max}}^2\,{\rm GeV}^2$ &
$~~~\theta_{\rm min}^{\rm cms}$~~~ \\ 
&  &  &  &  & \\ \hline
&  &  &  &  & \\ 
~~~~8.5  & ~0.065 & 0.00365 & 2.3  ~~~~~ & ~~~~~130  &   \,~~~ 15    \\ 
~~~~          &  & 0.00649 & 6.7  ~~~~~ & ~~~~~130  &   \,~~~ 26    \\ 
~~~~12   & ~0.065 & 0.00649 & 6.9  ~~~~~ & ~~~~~180  &   \,~~~ 22    \\ 
~~~~          &  & 0.01154 & 19   ~~~~~ & ~~~~~180  & ~~\,~   37    \\ 
~~~~          &  & 0.02052 & 44   ~~~~~ & ~~~~~180  & ~~\,~   59    \\ 	 
~~~~          &  & 0.03648 & 82   ~~~~~ & ~~~~~180  & ~~\,~   84    \\ 	 
~~~~     & ~0.175 & 0.00429 & 2.5  ~~~~~ & ~~~~~~69   &  \,~~~  22    \\ 	 
~~~~          &  & 0.00762 & 6.4  ~~~~~ & ~~~~~~69   &  \,~~~  36    \\ 
~~~~          &  & 0.01355 & 14   ~~~~~ & ~~~~~~69   &~~\,~    54    \\ 
~~~~          &  & 0.02410 & 26   ~~~~~ & ~~~~~~69   &~~\,~    75    \\ 
~~~~     & ~0.375 & 0.00356 & 1.1  ~~~~~ & ~~~~~~32   &  \,~~~  22    \\ 	 
~~~~          &  & 0.00632 & 2.7  ~~~~~ & ~~~~~~32   &  \,~~~  34    \\ 
~~~~          &  & 0.01125 & 5.5  ~~~~~ & ~~~~~~32   &  \,~~~  49    \\ 
~~~~     & ~0.65  & 0.00649 & 1.2  ~~~~~ & ~~~~~~18   &  \,~~~  29    \\ 	 
~~~~25   & ~0.065 & 0.01154 & 21   ~~~~~ & ~~~~~390  & ~~\,~   27    \\  
~~~~          &  & 0.02052 & 55   ~~~~~ & ~~~~~390  & ~~\,~   44    \\ 	 
~~~~          &  & 0.03648 & 120  ~~~~~ & ~~~~~390  & ~~\,~   68    \\ 	 
~~~~          &  & 0.06488 & 200  ~~~~~ & ~~~~~390  & ~~\,~   93    \\ 	 
~~~~     & ~0.175 & 0.00429 & 2.7  ~~~~~ & ~~~~~140  &   \,~~~ 16    \\  
~~~~          &  & 0.00762 & 7.4  ~~~~~ & ~~~~~140  &   \,~~~ 26    \\ 
~~~~          &  & 0.01355 & 18   ~~~~~ & ~~~~~140  & ~~\,~   42    \\ 	 
~~~~          &  & 0.02410 & 38   ~~~~~ & ~~~~~140  & ~~\,~   62    \\ 	 
~~~~          &  & 0.04285 & 62   ~~~~~ & ~~~~~140  & ~~\,~   83    \\ 	 
~~~~     & ~0.375 & 0.00356 & 1.3  ~~~~~ & ~~~~~~67   &  \,~~~  16    \\ 	 
~~~~          &  & 0.00632 & 3.3  ~~~~~ & ~~~~~~67   &  \,~~~  26    \\ 
~~~~          &  & 0.01125 & 7.5  ~~~~~ & ~~~~~~67   &  \,~~~  39    \\ 
~~~~          &  & 0.02000 & 14   ~~~~~ & ~~~~~~67   &~~\,~    55    \\ 
~~~~     & ~0.65  & 0.00649 & 1.5  ~~~~~ & ~~~~~~38   &  \,~~~  23    \\ 	 
~~~~          &  & 0.01154 & 3.1  ~~~~~ & ~~~~~~38   &  \,~~~  33    \\ 
~~~~50   & ~0.175 & 0.00762 & 8.2  ~~~~~ & ~~~~~290  &   \,~~~ 20    \\  
~~~~          &  & 0.01355 & 22   ~~~~~ & ~~~~~290  & ~~\,~   32    \\ 	 
~~~~          &  & 0.02410 & 51   ~~~~~ & ~~~~~290  & ~~\,~   50    \\ 	 
~~~~          &  & 0.04285 & 95   ~~~~~ & ~~~~~290  &  ~\,~~  71    \\ 	 
~~~~          &  & 0.07620 & 140  ~~~~~ & ~~~~~290  &   \,~~~ 90    \\ 	 
~~~~     & ~0.375 & 0.00356 & 1.4  ~~~~~ & ~~~~~130  &   \,~~~ 12    \\  
~~~~          &  & 0.00632 & 3.8  ~~~~~ & ~~~~~130  &   \,~~~ 19    \\ 
~~~~          &  & 0.01125 & 9.4  ~~~~~ & ~~~~~130  &   \,~~~ 31    \\ 
~~~~          &  & 0.02000 & 20   ~~~~~ & ~~~~~130  & ~~\,~   46    \\ 	 
~~~~          &  & 0.03556 & 35   ~~~~~ & ~~~~~130  & ~~\,~   62    \\ 	 
~~~~     & ~0.65  & 0.00649 & 1.9  ~~~~~ & ~~~~~~77   &  \,~~~  18    \\ 	 
~~~~          &  & 0.01154 & 4.2  ~~~~~ & ~~~~~~77   &  \,~~~  27    \\ 
~~~~          &  & 0.02052 & 7.9  ~~~~~ & ~~~~~~77   &  \,~~~  37    \\
\hline
\end{tabular}}
\caption[Struck quark virtuality constraints for Phillips'
data.]{{\it Virtuality constraints for diffractive scattering as a
function
of~$Q^2$,~$\beta$, and~$\xpom$ for the parameter range measured
in~\protect\cite{Phillips:1995jpp}. The constraints correspond to
the pseudo-rapidity cut of~$\eta_{\rm max}^{\rm exp}=1.8$ which was
used for data selection. Also shown is $k^2_{\rm max}$, which is the
maximum possible virtuality of the exchanged quark.}}
\label{table:phillipsvirtuality}
\end{table}

We have extended the original calculation to include large
pseudo-rapidity gap diffractive production of three or more partons,
which we term ``multi-jet'' production, due, e.g., to the diagrams
of~Fig.\,\ref{fig:3jets}. Here we consider the virtuality of the
exchanged parton coupling to the pomeron. In the boson-gluon fusion
case, for example, we are concerned with the virtuality of the
exchanged gluon. We find that the constraint on the virtuality of this
parton is slightly weaker than in the dijet case, but still find that
the large pseudo-rapidity gap selection cuts rule out
resolved-coupling contributions from these diagrams for most of the
data points corresponding
to~Table\,\ref{table:phillipsvirtuality}. The details of this
calculation are given in~Appendix\,A.2.

\section{Implications and Models of the Pseudo-Rapidity Gap Dependence}
\label{sect:ope}

The result of this analysis has been to show that, for a subclass of
the deep-inelastic diffractive events, the relevant colour-singlet
exchange process, pomeron exchange, must involve a direct coupling
to off-shell partons involved in the hard-scattering process. This is a
significant result because, at first sight, such a component is {\it
not} the usual parton contribution to inclusive deep-inelastic
scattering processes that is often used to interpret the data in terms
of a ``pomeron'' structure function. In particular, our results show
that the graphs involving $t$-channel colour non-singlet exchange
in the hard-scattering sub-process should not be
included in describing the deep-inelastic exclusive processes
involving strong pseudo-rapidity cuts. However, pseudo-rapidity cut
events exhibit scaling and form a significant part of the inclusive
deep-inelastic scattering events which {\it are} governed by the
operator product expansion (OPE). The latter ascribes the dominant
scaling contribution in deep-inelastic inclusive scattering to just
the colour exchange graphs which we argue must be absent from the
exclusive strong pseudo-rapidity cut events. These leading-twist graphs
have also been shown directly to dominate exclusive diffractive
processes, using an expansion in non-local operators
\cite{Hautmann:1998xn,Collins:1998sr}. However, it is not clear that
this analysis applies to the more exclusive pseudo-rapidity gap events of
interest here.

There are two possibilities for reconciling our results on the
importance of colour-singlet exchange with this expected
need for colour-exchange processes. The first arises because the
events being studied at HERA are at extremely small~$x_{\rm Bjorken}$,
in a region where the OPE, which is reliable when there is only one
large variable, may break down because~$Q^{2}$ and~$\nu$,
although both large, are not of the same order. We know that such a
breakdown must occur in the transition to forward scattering, which
is governed by~$t=0$ non-perturbative Regge-exchange processes. If
naive perturbation theory does fail, one is forced to adopt a more
phenomenological approach to the diffractive scattering of the struck
quarks off the proton target.  Donnachie and Landshoff have suggested that
diffractive quark-proton scattering occurs through a direct
coupling of a colourless pomeron to the
quark\,\cite{Donnachie:1987pu:Donnachie:1987xh}. In analogy with their
successful analysis of hadron-hadron diffractive
scattering 
\cite{Donnachie:1984hf:Donnachie:1984xq:Donnachie:1986iz:Donnachie:1992ny},
they assume that this process proceeds via vector exchange with a form
factor at the quark vertex to describe the dependence on the quark
virtuality. The latter is required to avoid power-law violations of
the observed scaling in deep-inelastic diffractive processes. In a similar
spirit, Vermaseren\,{\em et al.}~(VBLY)~\cite{Vermaseren:1996iy}
consider a
scalar effective exchange interaction to describe the same process.

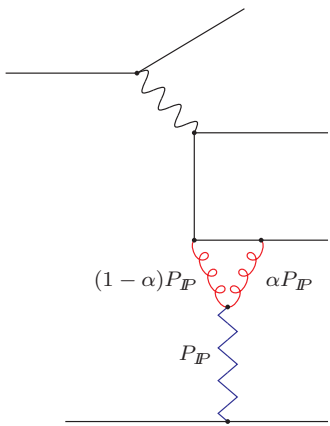
\begin{figure}[htp]
\begin{center} \begin{picture}(180,180)(-20,0) 
\SetScale{0.9}
\Line(5,148)(60,148)
\Line(60,148)(105,175)
\Line(84,123)(140,123)
\Line(84,78)(140,78)
\Line(84,123)(84,78)
\Line(30,2)(140,2)
\Photon(60,148)(84,123){4}{3.5}
\SetColor{Blue}
\ZigZag(98,50)(98,2){-4}{3.5}
\SetColor{Black}
\Text(38,55)[l]{\scriptsize $(1-\alpha)P_{\pom}$}
\Text(103,55)[l]{\scriptsize $\alpha P_{\pom}$}
\Text(70,27)[l]{\scriptsize $P_{\pom}$}
\SetColor{Red}
\Gluon(84,78)(98,50){-2.5}{3.5}
\Gluon(112,78)(98,50){2.5}{3.5}
\SetColor{Black}
\Vertex(60,148){1}
\Vertex(84,123){1}
\Vertex(84,78){1}
\Vertex(98,2){1}
\Vertex(98,50){1}
\Vertex(112,78){1}
\end{picture}
\caption[Two-gluon contribution to pomeron form factor.]{{\it Two-gluon
contribution to the pomeron form factor.}\label{fig:2glue}}
\end{center}
\end{figure}

A second possible way to resolve the apparent discrepancy between
the observation of scaling in the large pseudo-rapidity gap events and
the OPE does not involve abandoning the OPE for the values of~$x_{\rm
Bjorken}$ probed at HERA. This possibility is motivated by the
observation that
the~$x_{\rm Bjorken}$ and~$Q^{2}$ dependence of inclusive deep-inelastic
scattering is well fitted by a continuation of the usual
DGLAP analysis to small~$x_{\rm Bjorken}$. In this case, the large
pseudo-rapidity gap events, while not being directly described by the
OPE analysis, because of their exclusive nature, must obey
the constraints on the inclusive cross section of which they are a
part. In order to make connection with the usual QCD description of
deep-inelastic scattering, it is necessary to couple the pomeron to
near-mass-shell quarks and/or gluons. Given the discussion above, this
must be via colour-singlet exchange due to a diagram such as that in
Fig.\,\ref{fig:2glue}. How does such a diagram fit into the OPE
framework? At
first sight, it would seem that Fig.\,\ref{fig:2glue} corresponds to a
higher-twist operator contribution, because we cannot form a twist-two
operator from the four physical gluons involved in the square of the
scattering amplitude. This would mean that the associated structure
function should be suppressed relative to the twist-two contribution
by a factor~${\cal O}\left(\Lambda^{2}/Q^{2}\right)$, where~$\Lambda$ is
a hadronic scale, i.e., the associated structure function should not
scale, in disagreement
with the measured structure function. However, the identification of
the diagrams of Fig.\,\ref{fig:2glue} with higher-twist operators is
not correct if the right-hand gluon leaves the quark near its mass
shell. In this case, the hard-scattering process is that involving the
left-hand gluon only, and the diagram should be identified with the
leading twist-two gluon operator. The subsequent soft-scattering
process involving the right-hand gluon should properly be included in
the soft dressing that is implicitly present in any description of
deep-inelastic scattering in order to obtain a final state involving
colourless hadrons. The soft parton emission processes discussed in
Appendix\,A.2 also contribute to this dressing. In the fully inclusive
scattering case the sum of such dressing forms a complete sum and thus
drops out of the calculation of the total cross section.

In what follows we shall discuss both the above possibilities, and try to
determine which best describes the pseudo-rapidity gap events. As
mentioned above, the direct coupling has been variously described in
terms of a vector-like coupling to
quarks\,\cite{Donnachie:1987pu:Donnachie:1987xh,Ellis:1996cg}, and also
in terms of a model in which the pomeron has a point-like scalar coupling
to quarks and gluons\footnote{Here we consider only the dijet
component of the VBLY model.}\,\cite{Vermaseren:1996iy}.
The vector-like exchange model is motivated by the success of the
Donnachie-Landshoff (D-L) description of diffractive proton-proton and
proton-antiproton scattering in terms of vector exchange with Wu-Yang
couplings to the nucleon. They suggested that Pomeron exchange should
be treated as a~$C$-even vector-exchange process with~$\gamma_{\mu}$
couplings to the incident proton and the quark involved in the
deep-inelastic process as shown in Fig.\,\ref{fig:dijet}\footnote{Note that
the soft parton (gluon) emission processes discussed in Appendix\,A.2
do {\em not} fit into this description, because they correspond to the
scattering of a $C-$odd gluon (leading-twist) on the struck
quark.}. The expected Regge behaviour at large centre-of-mass energies is
put in by hand. In order to generate a cross section consistent with the
observed scaling, they introduced a {\em form~factor} at the quark
vertex of the form

\be \label{eq:dl}
f(k^2)=\frac{\Lambda^2}{\Lambda^2-k^2},
\ee

\noindent where~$k$ is the four momentum of the struck quark~($0\leq
|k^{2}|\leq Q^{2}/\beta$) in~Fig.\,\ref{fig:dijet}
and~$\Lambda = {\cal O}(\Lambda_{\rm QCD})$. This choice of form factor
allows
the pomeron to couple to off-shell partons, but results in the
dominant contribution to the diffractive cross section coming from
small~$k^{2}.$ This means that the model is very sensitive to strong
pseudo-rapidity cuts, and it was argued in~\cite{Ellis:1996cg}
that the cross section falls off too rapidly with
decreasing pseudo-rapidity to be able to explain the number of events
seen in the large pseudo-rapidity gap sample.

A variant of the model was considered in~\cite{Ellis:1996cg} (E-R),
where a harder form factor was chosen:

\be  \label{eq:er}
f(k^2)=\sqrt{\frac{\Lambda^2}{\Lambda^2-k^2}},
\ee

\noindent which leads to a prediction of additional logarithmic
scaling violations. Possible motivation for such a form comes from
underlying QCD diagrams, as shown in Fig.\,\ref{fig:2glue}. As
compared to the diagram of Fig.\,\ref{fig:dijet}, we have an extra
fermion propagator between the two gluon lines, leading to the harder
form factor of~(\ref{eq:er}). Such a choice is consistent with
measurements\,\cite{Adloff:1997sc} of the diffractive structure
function~$F_{2}^{D(3)}$. Since this form factor falls off more slowly
with $k^2$ than that of~(\ref{eq:dl}), the contribution from higher
virtuality states is enhanced. In fact, using~(\ref{eq:er}), one
finds a uniform contribution from all virtualities up to the
maximum\,\cite{Ellis:1996cg}.

\begin{figure}[htp]
\begin{center} \begin{picture}(300,95)(40,0)
\SetScale{0.5}
\Line(5,128)(60,128)
\Line(60,128)(105,155)
\Line(84,103)(130,103)
\Line(84,58)(130,58)
\Line(84,103)(84,58)
\Line(30,13)(130,13)
\Photon(60,128)(84,103){4}{3.5}
\Gluon(92,58)(92,13){-4}{4.5}
\Gluon(106,58)(106,13){-4}{4.5}
\Vertex(60,128){1}
\Vertex(84,103){1}
\Vertex(92,58){1}
\Vertex(106,58){1}
\Vertex(92,13){1}
\Vertex(106,13){1}

\Line(175,128)(230,128)
\Line(230,128)(275,155)
\Line(254,103)(300,103)
\Line(254,58)(300,58)
\Line(254,103)(254,58)
\Line(200,13)(300,13)
\Photon(230,128)(254,103){4}{3.5}
\Gluon(262,103)(262,13){-4}{8.5}
\Gluon(276,103)(276,13){-4}{8.5}
\Vertex(230,128){1}
\Vertex(254,103){1}
\Vertex(262,13){1}
\Vertex(276,13){1}
\Vertex(262,103){1}
\Vertex(276,103){1}

\Line(345,128)(400,128)
\Line(400,128)(445,155)
\Line(424,103)(470,103)
\Line(424,58)(470,58)
\Line(424,103)(424,58)
\Line(370,13)(470,13)
\Photon(400,128)(424,103){4}{3.5}
\Gluon(432,58)(432,13){-4}{4.5}
\Gluon(446,103)(446,13){-4}{8.5}
\Vertex(400,128){1}
\Vertex(426,103){1}
\Vertex(432,58){1}
\Vertex(446,103){1}
\Vertex(432,13){1}
\Vertex(446,13){1}

\Line(515,128)(570,128)
\Line(570,128)(615,155)
\Line(594,103)(640,103)
\Line(594,58)(640,58)
\Line(594,103)(594,58)
\Line(540,13)(640,13)
\Photon(570,128)(594,103){4}{3.5}
\Gluon(602,103)(602,13){-4}{8.5}
\Gluon(616,58)(616,13){-4}{4.5}
\Vertex(570,128){1}
\Vertex(594,103){1}
\Vertex(602,103){1}
\Vertex(616,58){1}
\Vertex(602,13){1}
\Vertex(616,13){1}
\end{picture}
\caption{{\it Two-gluon exchange graphs used to model pomeron
exchange.}\label{fig:coupleofgluons}}
\end{center}
\end{figure}
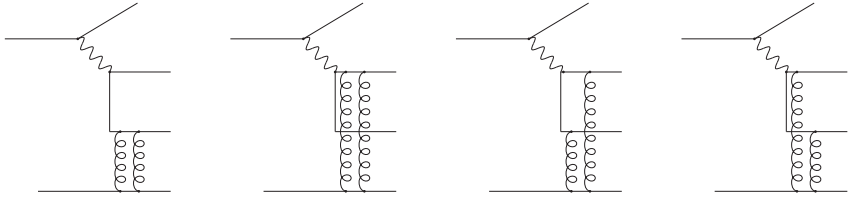

The third direct coupling model assumes effective scalar exchange
as treated in~\cite{Vermaseren:1996iy}. In this case, no form factor is
needed, as the cross section scales without it.

An alternative and possibly complementary approach to modelling the
direct (colour-singlet) coupling of the pomeron to the quarks is via
the colour singlet-component of multi-gluon exchange diagrams, in
which the gluons couple to the quarks involved in the hard-scattering
process. The simplest graphs are the two-gluon exchange processes
shown in Fig.\,\ref{fig:coupleofgluons}~\cite{multiglue}. Clearly,
this can only be an approximation to pomeron exchange, but it does
capture some of the important features, and does contribute to the
large pseudo-rapidity gap events. As is clear from our discussion
above, the direct coupling models {\em \`{a} la} Donnachie-Landshoff
contain a component of two-gluon exchange, but the latter has the
advantage of allowing for a non-local coupling of the pomeron to both
the struck quarks.

Of course these two interpretations need not be
distinct, in the sense that the two-gluon contribution of
Fig.\,\ref{fig:2glue} to the pomeron may be interpreted as a direct
coupling of the pomeron to quarks with a form factor. In this case, the
first two diagrams of Fig.\,\ref{fig:coupleofgluons} may be
interpreted as part of the direct contribution. However, gauge
invariance then requires that the remaining two diagrams of
Fig.\,\ref{fig:coupleofgluons} be included, and these cannot be
interpreted as a direct coupling of the pomeron to a single
quark. However, due to the extra hard propagator involved in these
terms, their contribution occurs only at higher twist. A consequence of
this is that, beyond leading twist, the direct coupling models at best
apply in a specific gauge, namely the gauge in which the contributions
of the diagrams of the type corresponding to the third and fourth
diagrams of Fig.\,\ref{fig:coupleofgluons} are minimized
\cite{Diehl:1995wz,Diehl:1998pd}.

On the other hand, as we have also discussed, a component of these
multi-gluon exchange graphs should be identified with the
leading-twist exchange. In the two-gluon case, the leading-twist
contribution corresponds to the imaginary part of the amplitude which
is indeed thought to dominate diffractive scattering. In this part,
the intermediate quark coupling to the right-hand gluon is on shell,
and this is the component that we argued has a leading-twist
component. For the case of more than two gluons, the projection onto
the imaginary part is not sufficient to avoid large quark propagators
appearing. The phase space associated with the relevant leading-twist
configuration is quite small for the large pseudo-rapidity gap
events. To see this, note that the pseudo-rapidity cut
requires~$2k.p>-k_{\rm min}^{2}.$ Thus, to keep the quark propagators
to the right of the left-hand gluon close to mass shell, the fraction,
1-$\alpha ,$ of the pomeron momentum carried by the left-hand gluon
must be large,~$\alpha\leq\Lambda ^{2}/2k\cdot
p\leq\Lambda^{2}/(-k_{\rm min}^{2})$, corresponding to a small phase
space factor for such effects. This suggests that either the two-gluon
graph dominates, in which case this constraint follows for the
dominant imaginary part, or there are a large number of such soft
gluons associated with the colour field of the proton remnant, so that
the sum over the ``dressing'' gluons compensates for the small phase
space available. The latter picture has been proposed by
Buchm\"{u}ller and Hebecker\,\cite{Buchmuller:1995qa}. In this
picture, the primary hard scattering is due to the usual one-gluon
exchange, followed by random soft dressing in the field of the proton
remnant. As discussed above, this picture is entirely consistent with
the OPE or related \cite{Hautmann:1998xn,Collins:1998sr} expansions,
and corresponds to keeping the leading-twist contribution. This
leading-twist interpretation of the pseudo-rapidity gap events does
not fit easily with the ``particle'' interpretation of the Pomeron as
a bound state with a significant two-gluon component, because there is
no obvious way that the soft gluon dressing should build up a pole in
the $t$ channel corresponding to the Pomeron singularity.

Note again that the fact that the pseudo-rapidity cut requires
colour-singlet exchange in the $t$-channel, combined with the
leading-twist constraint, requires that a single gluon carries the
bulk of the pomeron momentum. This is true also of the soft-parton
emission processes which are discussed in Appendix\,A.2, and which
{\em are} included in the leading-twist contribution. The requirement
that a single parton carries most of the Pomeron momentum is a purely
kinematical constraint, and does not imply that the same is true for
the full partonic structure of the Pomeron. However, in the
leading-twist model, the fact that the event rate requiring a
stringent pseudo-rapidity gap is comparable to the event rate without
this cut does suggest such a component is a significant part of the
partonic structure.

\section{Pseudo-rapidity Gap Tests}

\label{sect:tests}

We turn now to the comparison of these models with experiment. The
data we use are the large pseudo-rapidity cut events described above,
and Table\,\ref{table:phillipsvirtuality} shows the parameters of the
data set chosen, where the experimental cut on pseudo-rapidity
is~$\etamax =1.8$. From this Table, we can see that the virtuality
constraints rule out any contribution from dijet
production. Appendix\,A.2 shows that the multi-jet contribution has a
lower limit on the virtuality of the exchanged parton coupled to the
pomeron, so, as we can see from~Table\,\ref{table:phillipsvirtuality},
most of the data sample corresponds to regions where multi-jet events
are also not selected.

It is clearly of interest to try to
use these data to discriminate between the various models suggested 
to describe deep-inelastic diffractive scattering. Here we take a step in
this direction by computing the dependence of the deep-inelastic
diffractive scattering cross section and diffractive and pomeron
structure functions on the pseudo-rapidity cut for several of the
models discussed above. 

We first compute this dependence for
the case in which pomeron exchange is modelled by a vector~$C=+1$
exchange,
with a form factor at the quark vertex as discussed above. We also
consider the VBLY model, in which the pomeron is
assumed to have a scalar coupling to quarks.
The second class of model we consider is the leading-twist
Buchm\"{u}ller-Hebecker type model, in which a single gluon is responsible
for the hard scattering. In this case, the pseudo-rapidity gap dependence
is due to single-gluon exchange as in exclusive deep-inelastic
scattering. The second (and possibly further) gluon is then simply
there to provide soft dressing to produce colour-singlet final states,
and is assumed not to affect the pseudo-rapidity cut dependence.
Finally, we consider the case in which the pomeron is replaced by two
gluons, following the calculation of Diehl\,\cite{Diehl:1995wz}. We
will also comment at the end on the related two-gluon model
of\,\cite{Bartels:1998ea}.

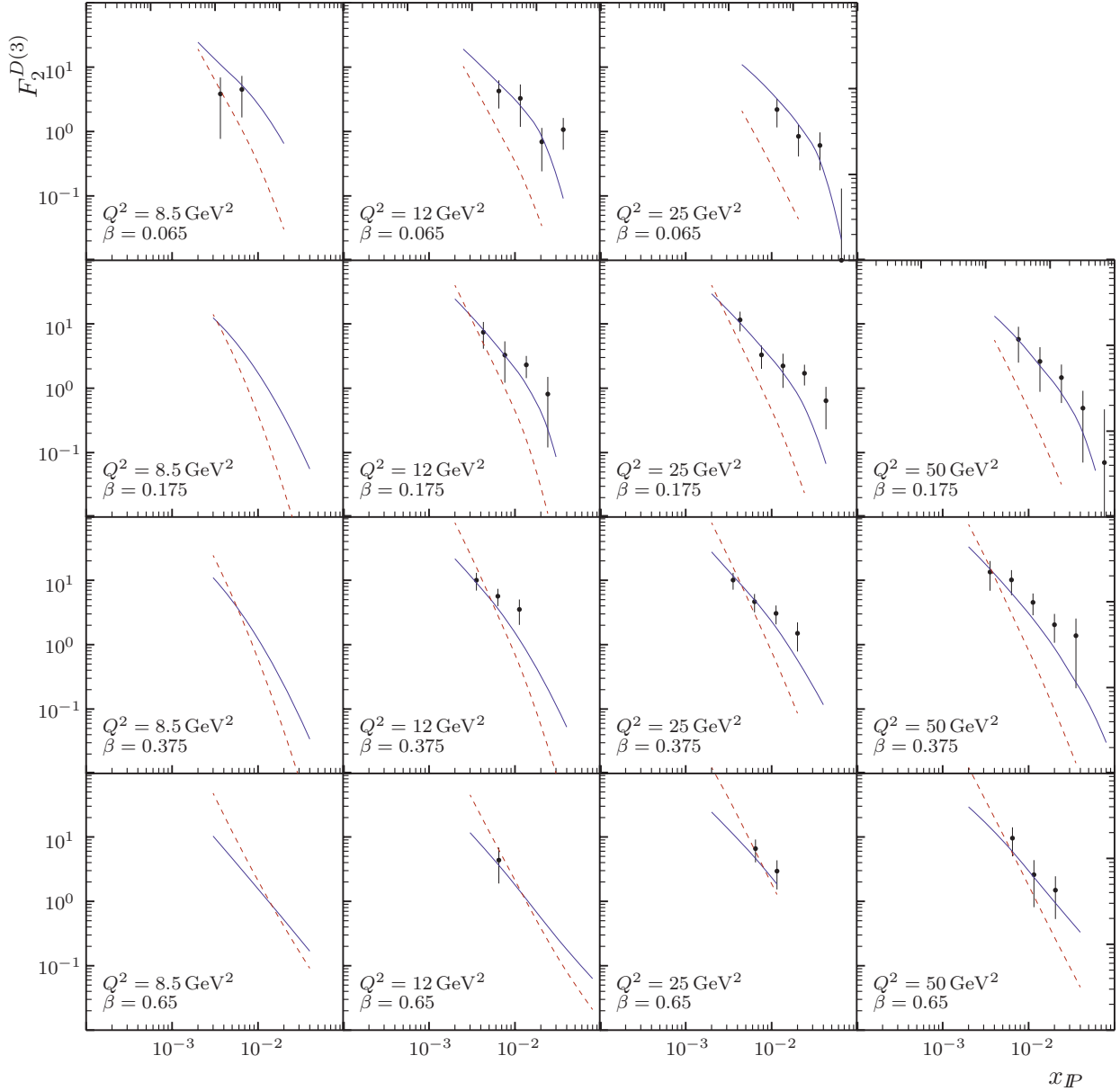
\begin{figure}[htp]
\begin{center} \begin{picture}(1.5,1.5)(0,0)
\SetScale{72}
\SetOffset(-200,-440)
\SetWidth{.001}
\LogAxis(0,0)(1.5,0)(3,.040,1,.0100)
\LogAxis(0,0)(0,1.5)(4,-.040,1,.0100)
 \SetColor{BlueViolet}
\Curve{(.7386,1.13)(.9515,0.88)(1.301,.46)}
 \SetColor{BrickRed}
\DashCurve{(.7386,1.38)(.9515,0.96)(1.301,.36)}{.03}
 \SetColor{Black}
\Text(37,-7)[]{\scriptsize $10^{-3}$}
\Text(74,-7)[]{\scriptsize $10^{-2}$}
\Text(-11,27)[]{\scriptsize $10^{-1}$}
\Text(-11,54)[]{\scriptsize $10^{0}$}
\Text(-11,80)[]{\scriptsize $10^{1}$}

\Text(7,20)[l]{\scriptsize $Q^2=8.5\,{\rm GeV}^2$}
\Text(7,11)[l]{\scriptsize $\beta=0.65$}

\SetOffset(-200,-332)
\SetWidth{.001}
\LogAxis(0,0)(1.5,0)(3,.040,1,.0100)
\LogAxis(0,0)(0,1.5)(4,-.040,1,.0100)
\SetColor{BlueViolet}
\Curve{(.7386,1.14)(.9515,.86)(1.301,.20)}
\SetColor{BrickRed}
\DashCurve{(.7386,1.27)(.9515,.79)(1.231,0)}{.03}
\SetColor{Black}
\Text(-11,27)[]{\scriptsize $10^{-1}$}
\Text(-11,54)[]{\scriptsize $10^{0}$}
\Text(-11,80)[]{\scriptsize $10^{1}$}

\Text(7,20)[l]{\scriptsize $Q^2=8.5\,{\rm GeV}^2$}
\Text(7,11)[l]{\scriptsize $\beta=0.375$}

\SetOffset(-200,-224)
\SetWidth{.001}
\LogAxis(0,0)(1.5,0)(3,.040,1,.0100)
\LogAxis(0,0)(0,1.5)(4,-.040,1,.0100)
\SetColor{BlueViolet}
\Curve{(.7386,1.16)(.9515,.91)(1.301,.28)}
\SetColor{BrickRed}
\DashCurve{(.7386,1.18)(.9515,.72)(1.199,0)}{.03}
\SetColor{Black}
\Text(-11,27)[]{\scriptsize $10^{-1}$}
\Text(-11,54)[]{\scriptsize $10^{0}$}
\Text(-11,80)[]{\scriptsize $10^{1}$}

\Text(7,20)[l]{\scriptsize $Q^2=8.5\,{\rm GeV}^2$}
\Text(7,11)[l]{\scriptsize $\beta=0.175$}

\SetOffset(-200,-116)
\SetWidth{.001}
\LogAxis(0,0)(1.5,0)(3,.040,1,.0100)
\LogAxis(0,1.5)(1.5,1.5)(4,-.040,1,.0100) %
\LogAxis(0,0)(0,1.5)(4,-.040,1,.0100)
\Vertex(.7811,.969){.014}
\Curve{(.7811,.707)(.7811,1.065)}
\Vertex(.9061,.995){.014}
\Curve{(.9061,.832)(.9061,1.074)}
\SetColor{BlueViolet}
\Curve{(.6505,1.27)(.7811,1.14)(.9061,1.02)(.9515,.97)(1.1505,.68)}
\SetColor{BrickRed}
\DashCurve{(.6505,1.23)(.7811,0.99)(.9061,0.76)(.9515,.67)(1.1505,.18)}{.03}
\SetColor{Black}

\Text(-11,27)[]{\scriptsize $10^{-1}$}
\Text(-11,54)[]{\scriptsize $10^{0}$}
\Text(-11,80)[]{\scriptsize $10^{1}$}

\Text(7,20)[l]{\scriptsize $Q^2=8.5\,{\rm GeV}^2$}
\Text(7,11)[l]{\scriptsize $\beta=0.065$}

\rText(-27,82)[][l]{\small $\fd$}

\SetOffset(-92,-440)
\SetWidth{.001}
\LogAxis(0,0)(1.5,0)(3,.040,1,.0100)
\LogAxis(0,0)(0,1.5)(4,-.040,1,.0100)
\Vertex(.9061,.990){.014}
\Curve{(.9061,.854)(.9061,1.063)}
\SetColor{BlueViolet}
\Curve{(.7386,1.15)(.9061,0.96)(1.301,.47)(1.4515,.30)}
\SetColor{BrickRed}
\DashCurve{(.7386,1.37)(.9061,1.05)(1.301,.34)(1.4515,.12)}{.03}
\SetColor{Black}
\Text(37,-7)[]{\scriptsize $10^{-3}$}
\Text(74,-7)[]{\scriptsize $10^{-2}$}
\Text(7,20)[l]{\scriptsize $Q^2=12\,{\rm GeV}^2$}
\Text(7,11)[l]{\scriptsize $\beta=0.65$}

\SetOffset(-92,-332)
\SetWidth{.001}
\LogAxis(0,0)(1.5,0)(3,.040,1,.0100)
\LogAxis(0,0)(0,1.5)(4,-.040,1,.0100)
\Vertex(.7757,1.125){.014}
\Curve{(.7757,1.065)(.7757,1.168)}
\Vertex(.9004,1.032){.014}
\Curve{(.9004,.974)(.9004,1.075)}
\Vertex(1.0256,.955){.014}
\Curve{(1.0256,.865)(1.0256,1.013)}
\SetColor{BlueViolet}
\Curve{(.6505,1.25)(.7757,1.11)(.9004,0.96)(1.0256,.78)(1.301,.27)}
\SetColor{BrickRed}
\DashCurve{(.6505,1.46)(.7757,1.20)(.9004,0.92)(1.0256,.63)(1.239,0)}{.03}
\SetColor{Black}
\Text(7,20)[l]{\scriptsize $Q^2=12\,{\rm GeV}^2$}
\Text(7,11)[l]{\scriptsize $\beta=0.375$}

\SetOffset(-92,-224)
\SetWidth{.001}
\LogAxis(0,0)(1.5,0)(3,.040,1,.0100)
\LogAxis(0,0)(0,1.5)(4,-.040,1,.0100)
\Vertex(.8162,1.076){.014}
\Curve{(.8162,.980)(.8162,1.136)}
\Vertex(.9410,.943){.014}
\Curve{(.9410,.782)(.9410,1.023)}
\Vertex(1.0660,.886){.014}
\Curve{(1.0660,.809)(1.0660,.938)}
\Vertex(1.1910,.716){.014}
\Curve{(1.1910,.405)(1.1910,.816)}
\SetColor{BlueViolet}
\Curve{(.6505,1.27)(.8162,1.09)(.9410,.94)(1.066,.77)(1.191,.51)(1.239,.35)}
\SetColor{BrickRed}
\DashCurve{(.6505,1.35)(.8162,1.01)(.9410,.75)(1.066,.45)(1.191,.02)}{.03}
\SetColor{Black}

\Text(7,20)[l]{\scriptsize $Q^2=12\,{\rm GeV}^2$}
\Text(7,11)[l]{\scriptsize $\beta=0.175$}

\SetOffset(-92,-116)
\SetWidth{.001}
\LogAxis(0,0)(1.5,0)(3,.040,1,.0100)
\LogAxis(0,1.5)(1.5,1.5)(4,-.040,1,.0100) %
\LogAxis(0,0)(0,1.5)(4,-.040,1,.0100)
\Vertex(.9061,.986){.014}
\Curve{(.9061,.884)(.9061,1.048)}
\Vertex(1.0311,.943){.014}
\Curve{(1.0311,.777)(1.0311,1.024)}
\Vertex(1.1561,.690){.014}
\Curve{(1.1561,.518)(1.1561,.771)}
\Vertex(1.2810,.761){.014}
\Curve{(1.2810,.644)(1.2810,.829)}
\SetColor{BlueViolet}
\Curve{(.6990,1.23)(.9061,1.03)(1.0311,.90)(1.1561,.72)(1.2810,.36)}
\SetColor{BrickRed}
\DashCurve{(.6990,1.13)(.9061,0.75)(1.0311,.51)(1.1561,.20)}{.03}
\SetColor{Black}
\Text(7,20)[l]{\scriptsize $Q^2=12\,{\rm GeV}^2$}
\Text(7,11)[l]{\scriptsize $\beta=0.065$}

\SetOffset(16,-440)
\SetWidth{.001}
\LogAxis(0,0)(1.5,0)(3,.040,1,.0100)
\LogAxis(0,0)(0,1.5)(4,-.040,1,.0100)
\Vertex(.9061,1.057){.014}
\Curve{(.9061,.977)(.9061,1.110)}
\Vertex(1.0311,.926){.014}
\Curve{(1.0311,.819)(1.0311,.989)}
\SetColor{BlueViolet}
\Curve{(.6505,1.27)(.7386,1.18)(.9061,1.00)(1.031,.85)}
\SetColor{BrickRed}
\DashCurve{(.6505,1.53)(.7386,1.37)(.9061,1.04)(1.031,.79)}{.03}
\SetColor{Black}
\Text(37,-7)[]{\scriptsize $10^{-3}$}
\Text(74,-7)[]{\scriptsize $10^{-2}$}
\Text(7,20)[l]{\scriptsize $Q^2=25\,{\rm GeV}^2$}
\Text(7,11)[l]{\scriptsize $\beta=0.65$}

\SetOffset(16,-332)
\SetWidth{.001}
\LogAxis(0,0)(1.5,0)(3,.040,1,.0100)
\LogAxis(0,0)(0,1.5)(4,-.040,1,.0100)
\Vertex(.7757,1.126){.014}
\Curve{(.7757,1.070)(.7757,1.168)}
\Vertex(.9004,1.000 ){.014}
\Curve{(.9004,0.938)(.9004,1.045)}
\Vertex(1.0256,0.932){.014}
\Curve{(1.0256,.868 )(1.0256,.978 )}
\Vertex(1.1505,.816 ){.014}
\Curve{(1.1505,.710 )(1.1505,.880 )}
\SetColor{BlueViolet}
\Curve{(.6505,1.29)(.7757,1.15)(.9004,1.01)(1.0256,.85)(1.1505,.66)(1.301,.40)}
\SetColor{BrickRed}
\DashCurve{(.6505,1.46)(.7757,1.20)(.9004,0.93)(1.0256,.65)(1.1505,.35)}{.03}
\SetColor{Black}
\Text(7,20)[l]{\scriptsize $Q^2=25\,{\rm GeV}^2$}
\Text(7,11)[l]{\scriptsize $\beta=0.375$}

\SetOffset(16,-224)
\SetWidth{.001}
\LogAxis(0,0)(1.5,0)(3,.040,1,.0100)
\LogAxis(0,0)(0,1.5)(4,-.040,1,.0100)
\Vertex(.8162,1.148){.014}
\Curve{(.8162,1.081)(.8162,1.196)}
\Vertex(.9410,.944){.014}
\Curve{(.9410,.863)(.9410,.998)}
\Vertex(1.0660,.880){.014}
\Curve{(1.0660,.752)(1.0660,.951)}
\Vertex(1.1910,.837){.014}
\Curve{(1.1910,.766)(1.1910,.887)}
\Vertex(1.3160,.677){.014}
\Curve{(1.3160,.511)(1.3160,.758)}
\SetColor{BlueViolet}
\Curve{(.6505,1.30)(0.8162,1.13)(0.9410,.99)(1.0660,.84)(1.191,.64)(1.3160,.31)}
\SetColor{BrickRed}
\DashCurve{(.6505,1.35)(0.8162,1.01)(0.9410,.75)(1.0660,.47)(1.191,.14)}{.03}
\SetColor{Black}
\Text(7,20)[l]{\scriptsize $Q^2=25\,{\rm GeV}^2$}
\Text(7,11)[l]{\scriptsize $\beta=0.175$}

\SetOffset(16,-116)
\SetWidth{.001}
\LogAxis(0,0)(1.5,0)(3,.040,1,.0100)
\LogAxis(0,1.5)(1.5,1.5)(4,-.040,1,.0100) %
\LogAxis(0,0)(0,1.5)(4,-.040,1,.0100)
\Vertex(1.0311,.878){.014}
\Curve{(1.0311,.774)(1.0311,.940)}
\Vertex(1.1561,.722){.014}
\Curve{(1.1561,.605)(1.1561,.789)}
\Vertex(1.2810,.669){.014}
\Curve{(1.2810,.524)(1.2810,.745)}
\Vertex(1.4061,0){.014}
\Curve{(1.4061,0)(1.4061,.418)}
\SetColor{BlueViolet}
\Curve{(.8266,1.14)(1.0311,.94)(1.1561,.79)(1.2810,.58)(1.3495,.36)(1.4061,.12)}
\SetColor{BrickRed}
\DashCurve{(.8266,0.87)(1.0311,.49)(1.1561,.24)}{.03}
\SetColor{Black}
\Text(7,20)[l]{\scriptsize $Q^2=25\,{\rm GeV}^2$}
\Text(7,11)[l]{\scriptsize $\beta=0.065$}

\SetOffset(124,-440)
\SetWidth{.001}
\LogAxis(0,0)(1.5,0)(3,.040,1,.0100)
\LogAxis(0,0)(0,1.5)(4,-.040,1,.0100)
\LogAxis(1.5,0)(1.5,1.5)(3,.040,0,.0100) %
\Vertex(.9061,1.118){.014}
\Curve{(.9061,1.012)(.9061,1.181)}
\Vertex(1.0311,.905){.014}
\Curve{(1.0311,.716)(1.0311,.990)}
\Vertex(1.1561,.815){.014}
\Curve{(1.1561,.647)(1.1561,.896)}
\SetColor{BlueViolet}
\Curve{(.6505,1.30)(.9061,1.04)(1.0311,.89)(1.1561,.74)(1.301,.57)}
\SetColor{BrickRed}
\DashCurve{(.6505,1.53)(.9061,1.03)(1.0311,.78)(1.1561,.53)(1.301,.25)}{.03}
\SetColor{Black}
\Text(37,-7)[]{\scriptsize $10^{-3}$}
\Text(74,-7)[]{\scriptsize $10^{-2}$}
\Text(7,20)[l]{\scriptsize $Q^2=50\,{\rm GeV}^2$}
\Text(7,11)[l]{\scriptsize $\beta=0.65$}

\Text(88,-21)[]{\small $x_{I\!\!P}$}

\SetOffset(124,-332)
\SetWidth{.001}
\LogAxis(0,0)(1.5,0)(3,.040,1,.0100)
\LogAxis(0,0)(0,1.5)(4,-.040,1,.0100)
\LogAxis(1.5,0)(1.5,1.5)(3,.040,0,.0100) %
\Vertex(.7757,1.173){.014}
\Curve{(.7757,1.064)(.7757,1.237)}
\Vertex(.9004,1.127){.014}
\Curve{(.9004,1.038)(.9004,1.184)}
\Vertex(1.0256,.996){.014}
\Curve{(1.0256,.921)(1.0256,1.048)}
\Vertex(1.1505,.866){.014}
\Curve{(1.1505,.761)(1.1505,.929)}
\Vertex(1.2755,.802){.014}
\Curve{(1.2755,.496)(1.2755,.902)}
\SetColor{BlueViolet}
\Curve{(.6505,1.32)(.7757,1.19)(0.9004,1.05)(1.0256,.90)(1.1505,.73)(1.2755,.53)(1.301,.49)(1.4515,.18)}
\SetColor{BrickRed}
\DashCurve{(.6505,1.45)(.7757,1.20)(0.9004,0.93)(1.0256,.66)(1.1505,.37)(1.2755,.06)}{.03}
\SetColor{Black}

\Text(7,20)[l]{\scriptsize $Q^2=50\,{\rm GeV}^2$}
\Text(7,11)[l]{\scriptsize $\beta=0.375$}

\SetOffset(124,-224)
\SetWidth{.001}
\LogAxis(0,0)(1.5,0)(3,.040,1,.0100)
\LogAxis(0,0)(0,1.5)(4,-.040,1,.0100)
\LogAxis(1.5,0)(1.5,1.5)(3,.040,0,.0100) %
\Vertex(.9410,1.035){.014}
\Curve{(.9410,.899)(.9410,1.108)}
\Vertex(1.0660,.906){.014}
\Curve{(1.0660,.729)(1.0660,.989)}
\Vertex(1.1910,.812){.014}
\Curve{(1.1910,.664)(1.1910,.888)}
\Vertex(1.3160,.634){.014}
\Curve{(1.3160,.317)(1.3160,.735)}
\Vertex(1.4410,.317){.014}
\Curve{(1.4410,0)(1.4410,.627)}
\SetColor{BlueViolet}
\Curve{(.8010,1.17)(0.9410,1.03)(1.0660,.88)(1.1910,.72)(1.3160,.49)(1.3891,.27)}
\SetColor{BrickRed}
\DashCurve{(.8010,1.03)(0.9410,0.75)(1.0660,.48)(1.1910,.19)}{.03}
\SetColor{Black}
\Text(7,20)[l]{\scriptsize $Q^2=50\,{\rm GeV}^2$}
\Text(7,11)[l]{\scriptsize $\beta=0.175$}

\SetOffset(124,-116)
\SetWidth{.001}
\LogAxis(0,0)(1.5,0)(4,-.040,1,.0100)
\LogAxis(0,0)(0,1.5)(3,.040,1,.0100)
\end{picture}
\vspace*{6.3in}
\caption[Structure function $\fd$ for the E-R and D-L form factor models.]
{{\it The E-R and D-L form-factor
models are fitted to data for $F_{2}^{D(3)}$ with a virtuality cut.
The data sample corresponds to a pseudo-rapidity cut at
hadron level of $\eta_{\rm max}=1.8$, the pomeron intercept is here
assumed to be~$\alpha_{\pom}(0)=1.08$, and in the form factors of
(\ref{eq:dl}) (D-L) and (\ref{eq:er}) (E-R) the parameter
$\Lambda$ is chosen to be 0.2\,GeV. The points correspond to data from
\cite{Phillips:1995jpp} for which~$-k^2_{\rm min}>\,1\,{\rm
GeV}^2$, and the statistical and systematic errors
have been added in quadrature.
The solid line is the prediction from the E-R model
with the overall normalization determined by a fit to the data points. The
dashed line is the D-L model with the same
normalization procedure.}\label{fig:erdl108multigraph}}
\end{center}
\end{figure}

We start with the direct-coupling models with vector-like coupling. In
Fig.\,\ref{fig:erdl108multigraph} we show the fit to the diffractive
structure function of~(\ref{eq:sfn}) for the pseudo-rapidity cut data
of~\cite{Phillips:1995jpp}. The solid points are experimental
data, the solid line represents the E-R model, and the dotted
line corresponds to the D-L model. The overall normalization is the
sole parameter of each model, and is determined by a fit to all the
data points. The plots shown here correspond to a pomeron intercept
of~$\alpha_{I\!\!P}(0)=1.08$, corresponding to the soft pomeron
intercept of hadron-hadron elastic and diffractive
scattering\,\cite{Donnachie:1984hf:Donnachie:1984xq:%
Donnachie:1986iz:Donnachie:1992ny}. However, as is discussed further
below, we have also obtained similar fits for other choices
of the pomeron intercept.
From~Fig.\,\ref{fig:erdl108multigraph} we can see that the E-R model
provides a rather good fit, showing that the harder form factor is an
improvement over that used in the D-L model. This point is made in a
more quantitative way in Table\,\ref{table:summary}, where the $\chi^2$
for the various fits are given.

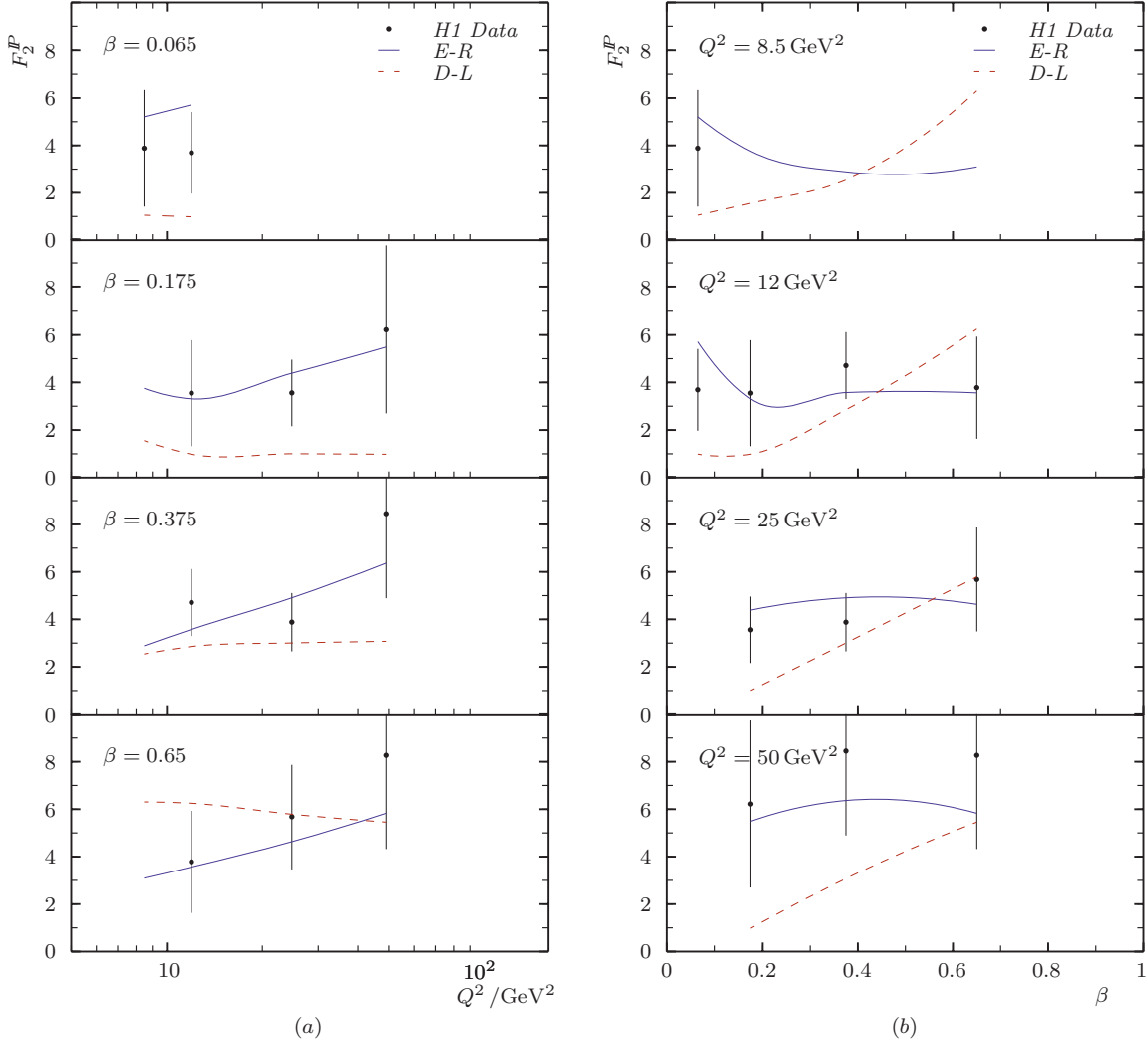
\begin{figure}[htp]
\begin{center} \begin{picture}(2.5,1.25)(0,-20)
\SetScale{72}  
\SetOffset(-200,-440)
\SetWidth{.001}
\LinAxis(0,0)(0,1.25)(5,2,-.040,0,.0100)
\LogAxis(0,0)(2.5,0)(1.5,.040,5,.0100)
\SetColor{Black}
\SetScale{1}
\Vertex(45.4,34.02){1}
\Curve{(45.4,14.67)(45.4,53.37)}
\Vertex(83.4,51.12){1}
\Curve{(83.4,31.14)(83.4,70.83)}
\Vertex(119,74.43){1}
\Curve{(119,38.88)(119,90)}
\SetColor{BlueViolet}
\Curve{(27.5,27.84)(45.4,32.04)(83.4,41.68)(119,52.45)}
\SetColor{BrickRed}
\DashCurve{(27.5,56.74)(45.4,56.25)(83.4,52.09)(119,49.09)}{3}
\SetScale{72}
\SetColor{Black}
\Text(37,-7)[]{\scriptsize $10$}
\Text(155,-7)[]{\scriptsize $10^{2}$}
\Text(165,-17)[]{\scriptsize $Q^2\,/{\rm GeV}^2$}
\Text(155,-7)[]{\scriptsize $10^{2}$}

\Text(-8,0)[]{\scriptsize $0$}
\Text(-8,18)[]{\scriptsize $2$}
\Text(-8,36)[]{\scriptsize $4$}
\Text(-8,54)[]{\scriptsize $6$}
\Text(-8,72)[]{\scriptsize $8$}
\Text(90,-29)[]{\scriptsize $(a)$}

\Text(12,74)[l]{\scriptsize $\beta=0.65$}
\LinAxis(2.5,0)(2.5,1.25)(5,2,.040,0,.0100)

\SetOffset(-200,-350)
\SetWidth{.001}
\LinAxis(0,0)(0,1.25)(5,2,-.040,0,.0100)
\LogAxis(0,0)(2.5,0)(1.5,.040,5,.0100)
\SetScale{1}
\Vertex(45.4,42.39){1}
\Curve{(45.4,29.7)(45.4,55.08)}
\Vertex(83.4,34.94){1}
\Curve{(83.4,23.85)(83.4,45.99)}
\Vertex(119,76.05){1}
\Curve{(119,44.01)(119,90)}
\SetColor{BlueViolet}
\Curve{(27.5,25.96)(45.4,32.17)(83.4,44.17)(119,57.29)}
\SetColor{BrickRed}
\DashCurve{(27.5,22.85)(45.4,25.70)(83.4,27.05)(119,27.67)}{3}
\SetColor{Black}
\SetScale{72}

\Text(-8,0)[]{\scriptsize $0$}
\Text(-8,18)[]{\scriptsize $2$}
\Text(-8,36)[]{\scriptsize $4$}
\Text(-8,54)[]{\scriptsize $6$}
\Text(-8,72)[]{\scriptsize $8$}

\Text(12,74)[l]{\scriptsize $\beta=0.375$}

\LinAxis(2.5,0)(2.5,1.25)(5,2,.040,0,.0100)

\SetOffset(-200,-260)
\SetWidth{.001}
\LinAxis(0,0)(0,1.25)(5,2,-.040,0,.0100)
\LogAxis(0,0)(2.5,0)(1.5,.040,5,.0100)
\SetScale{1}
\Vertex(45.4,31.95){1}
\Curve{(45.4,11.88)(45.4,52.02)}
\Vertex(83.4,32.04){1}
\Curve{(83.4,19.44)(83.4,44.64)}
\Vertex(119,55.98){1}
\Curve{(119,24.3)(119,87.66)}
\SetColor{BlueViolet}
\Curve{(27.5,33.78)(45.4,29.79)(83.4,39.50)(119,49.41)}
\SetColor{BrickRed}
\DashCurve{(27.5,14.01)(45.4,8.860)(83.4,9.020)(119,8.820)}{3}
\SetColor{Black}
\SetScale{72}
 
\Text(-8,0)[]{\scriptsize $0$}
\Text(-8,18)[]{\scriptsize $2$}
\Text(-8,36)[]{\scriptsize $4$}
\Text(-8,54)[]{\scriptsize $6$}
\Text(-8,72)[]{\scriptsize $8$}
 
\Text(12,74)[l]{\scriptsize $\beta=0.175$}
\LinAxis(2.5,0)(2.5,1.25)(5,2,.040,0,.0100)
 
\SetOffset(-200,-170)
\SetWidth{.001}
\LinAxis(0,0)(0,1.25)(5,2,-.040,0,.0100)
\LogAxis(0,0)(2.5,0)(1.5,.040,5,.0100)
\SetScale{1}
\Vertex(120,80){1}
\SetColor{BlueViolet}
\Curve{(116,72)(124,72)}
\SetColor{BrickRed}
\DashCurve{(116,64)(124,64)}{3}
\SetColor{Black}
\Vertex(27.5,34.92){1}
\Curve{(27.5,12.78)(27.5,57.06)}
\Vertex(45.4,33.21){1}
\Curve{(45.4,17.73)(45.4,48.69)}
\SetColor{BlueViolet}
\Curve{(27.5,46.86)(45.4,51.42)}
\SetColor{BrickRed}
\DashCurve{(27.5,9.490)(45.4,8.890)}{3}
\SetColor{Black}
\SetScale{72}
 
\Text(-8,0)[]{\scriptsize $0$}
\Text(-8,18)[]{\scriptsize $2$}
\Text(-8,36)[]{\scriptsize $4$}
\Text(-8,54)[]{\scriptsize $6$}
\Text(-8,72)[]{\scriptsize $8$}
\rText(-21,72)[][l]{\scriptsize $F_{2}^{\pom}$}
 
\Text(137,80)[l]{\scriptsize\em H1 Data}
\Text(137,72)[l]{\scriptsize\em E-R}
\Text(137,64)[l]{\scriptsize\em D-L}
 
\Text(12,74)[l]{\scriptsize $\beta=0.065$}
 
\LinAxis(2.5,0)(2.5,1.25)(5,2,.040,0,.0100)
\LogAxis(0,1.25)(2.5,1.25)(1.5,-.040,5,.0100)
 
\SetOffset(26,-440)
\SetWidth{.001}
\LinAxis(0,0)(0,1.25)(5,2,-.040,0,.0100)
\LinAxis(0,0)(2.5,0)(5,2,.040,0,.0100)
\SetScale{1}
\Vertex(31.5,55.98){1}
\Curve{(31.5,24.3)(31.5,87.66)}
\Vertex(67.5,76.05){1}
\Curve{(67.5,44.01)(67.5,90)}
\Vertex(117,74.43){1}
\Curve{(117,38.88)(117,90)}
\SetColor{BlueViolet}
\Curve{(31.5,49.41)(67.5,57.29)(117,52.45)}
\SetColor{BrickRed}
\DashCurve{(31.5,8.820)(67.5,27.67)(117,49.09)}{3}
\SetScale{72}
\SetColor{Black}
\Text(0,-7)[]{\scriptsize $0$}
\Text(36,-7)[]{\scriptsize $0.2$}
\Text(72,-7)[]{\scriptsize $0.4$}
\Text(108,-7)[]{\scriptsize $0.6$}
\Text(144,-7)[]{\scriptsize $0.8$}
\Text(180,-7)[]{\scriptsize $1$}
\Text(165,-17)[]{\scriptsize $\beta$}
\Text(90,-29)[]{\scriptsize $(b)$}
 
\Text(-8,0)[]{\scriptsize $0$}
\Text(-8,18)[]{\scriptsize $2$}
\Text(-8,36)[]{\scriptsize $4$}
\Text(-8,54)[]{\scriptsize $6$}
\Text(-8,72)[]{\scriptsize $8$}
 
\Text(12,74)[l]{\scriptsize $Q^2=50\,{\rm GeV}^2$}
\LinAxis(2.5,0)(2.5,1.25)(5,2,.040,0,.0100)

\SetOffset(26,-350)
\SetWidth{.001}
\LinAxis(0,0)(0,1.25)(5,2,-.040,0,.0100)
\LinAxis(0,0)(2.5,0)(5,2,.040,0,.0100)
\SetScale{1}
\Vertex(31.5,32.04){1}
\Curve{(31.5,19.44)(31.5,44.64)}
\Vertex(67.5,34.92){1}
\Curve{(67.5,23.85)(67.5,45.99)}
\Vertex(117,51.12){1}
\Curve{(117,31.41)(117,70.83)}
\SetColor{BlueViolet}
\Curve{(31.5,39.50)(67.5,44.17)(117,41.68)}
\SetColor{BrickRed}
\DashCurve{(31.5,9.020)(67.5,27.05)(117,52.09)}{3}
\SetColor{Black}
\SetScale{72}
 
\Text(-8,0)[]{\scriptsize $0$}
\Text(-8,18)[]{\scriptsize $2$}
\Text(-8,36)[]{\scriptsize $4$}
\Text(-8,54)[]{\scriptsize $6$}
\Text(-8,72)[]{\scriptsize $8$}
 
\Text(12,74)[l]{\scriptsize $Q^2=25\,{\rm GeV}^2$}
\LinAxis(2.5,0)(2.5,1.25)(5,2,.040,0,.0100)

\SetOffset(26,-260)
\SetWidth{.001}
\LinAxis(0,0)(0,1.25)(5,2,-.040,0,.0100)
\LinAxis(0,0)(2.5,0)(5,2,.040,0,.0100)
\SetScale{1}
\Vertex(11.7,33.21){1}
\Curve{(11.7,17.73)(11.7,48.69)}
\Vertex(31.5,31.95){1}
\Curve{(31.5,11.88)(31.5,52.02)}
\Vertex(67.5,42.39){1}
\Curve{(67.5,29.7)(67.5,55.08)}
\Vertex(117,34.02){1}
\Curve{(117,14.67)(117,53.37)}
\SetColor{BlueViolet}
\Curve{(11.7,51.42)(31.5,29.79)(67.5,32.17)(117,32.04)}
\SetColor{BrickRed}
\DashCurve{(11.7,8.890)(31.5,8.860)(67.5,25.70)(117,56.25)}{3}
\SetColor{Black}
\SetScale{72}
 
\Text(-8,0)[]{\scriptsize $0$}
\Text(-8,18)[]{\scriptsize $2$}
\Text(-8,36)[]{\scriptsize $4$}
\Text(-8,54)[]{\scriptsize $6$}
\Text(-8,72)[]{\scriptsize $8$}
 
\Text(12,74)[l]{\scriptsize $Q^2=12\,{\rm GeV}^2$}
\LinAxis(2.5,0)(2.5,1.25)(5,2,.040,0,.0100)

\SetOffset(26,-170)
\SetWidth{.001}
\LinAxis(0,0)(0,1.25)(5,2,-.040,0,.0100)
\LinAxis(0,0)(2.5,0)(5,2,.040,0,.0100)
\SetScale{1}
 
\Vertex(120,80){1}
\SetColor{BlueViolet}
\Curve{(116,72)(124,72)}
\SetColor{BrickRed}
\DashCurve{(116,64)(124,64)}{3}
\SetColor{Black}
\Vertex(11.7,34.92){1}
\Curve{(11.7,12.78)(11.7,57.06)}
\SetColor{BlueViolet}
\Curve{(11.7,46.86)(31.5,33.78)(67.5,25.96)(117,27.84)}
\SetColor{BrickRed}
\DashCurve{(11.7,9.490)(31.5,14.01)(67.5,22.85)(117,56.74)}{3}
\SetColor{Black}
\SetScale{72}
 
\Text(-8,0)[]{\scriptsize $0$}
\Text(-8,18)[]{\scriptsize $2$}
\Text(-8,36)[]{\scriptsize $4$}
\Text(-8,54)[]{\scriptsize $6$}
\Text(-8,72)[]{\scriptsize $8$}
\rText(-21,72)[][l]{\scriptsize $F_{2}^{\pom}$}
 
\Text(137,80)[l]{\scriptsize\em H1 Data}
\Text(137,72)[l]{\scriptsize\em E-R}
\Text(137,64)[l]{\scriptsize\em D-L}
 
\Text(12,74)[l]{\scriptsize $Q^2=8.5\,{\rm GeV}^2$}
\LinAxis(2.5,0)(2.5,1.25)(5,2,.040,0,.0100)
\LinAxis(0,1.25)(2.5,1.25)(5,2,-.040,0,.0100)

\end{picture}
\vspace*{6.2in}
\caption[Structure function $\ftwopom$ for the E-R and D-L form factor
models.]  {{\it The E-R and D-L form factor models are fitted to the
virtuality cut data~\cite{Phillips:1995jpp} for the structure
function $\ftwopom$, with parameters as in
Fig.\,\ref{fig:erdl108multigraph}. The solid line is the prediction
from the E-R model as determined in Fig.\,\ref{fig:erdl108multigraph},
and the dashed line is the D-L model.}\label{fig:erdlf2d2108}}
\end{center}
\end{figure}

Fig.\,\ref{fig:erdlf2d2108}a shows the pomeron structure
function,~$F_2^{I\!\!P}$, of (\ref{eq:pomsfn}), up to the
normalization determined above, as a function of~$Q^2$ at fixed
$\beta$, and Fig.\,\ref{fig:erdlf2d2108}b shows the~$\beta$
dependence of $F_{2}^{I\!\!P}$ at fixed $Q^2$.
Na\"{i}vely, one can determine the approximate behaviour of the
diffractive structure function as a function of $\beta$, at fixed
$\qsq$ and $\xpom$, by looking at the behaviour\footnote{In the case
that the diffractive structure function factorizes to give a pomeron
structure function, this is the dependence of $\ftwopom$ at fixed
$\qsq$.}  of $\fd$ at small virtuality, $k^2$. Then, for example,
in the case of the D-L model one would expect to find that

\be
\ftwopom\sim\beta(1-\beta).
\ee

\noindent
Further, one can show that in the case of large virtuality constraints
($\kmin>1\,{\rm GeV}^2$), the form of $\ftwopom$ is modified to

\be
\ftwopom\sim\beta(1-\beta)^2.
\ee

\noindent
However, there is an additional contribution to the $\beta$ dependence
in experiments with a large virtuality constraint,
arising from the lower limit
of the $k^2$ phase space integral. We can see from
Table\,\ref{table:phillipsvirtuality} that, at fixed $\qsq$ and
$\xpom$, $\kmin$ decreases approximately linearly with increasing
$\beta$. Hence, for the D-L model, in which the form factor introduces
an approximately $1/k^4$ dependence into the cross section, we would
expect to see a rather strong suppression in the ``pomeron structure
function'' at small $\beta$. This is clearly observed in the dotted
curve of Fig.\,\ref{fig:erdlf2d2108}b.

The harder form factor model of E-R has an approximately $1/k^2$
dependence introduced by the form factor, so such a strong suppression
is not expected. Furthermore, this choice of form factor leads to a
$\ln(\qsq/\mu^2)$ behaviour in the diffractive cross section, which
reflects contributions from the whole range of $k^2$, and hence the
small $k^2$ approximation is expected to be relatively poor. In
Fig.\,\ref{fig:erdlf2d2108}b we see that, even for data with a
relatively large rapidity cut, the E-R model predicts a rather flat
$\beta$ distribution of $\ftwopom$, in better agreement with the
experimental data. Further, as discussed earlier, as a result of the
large pseudo-rapidity gap cuts, we expect a further $\xpom$ dependence
from the lower limit of the phase-space integral. This means that a
``pomeron structure function'' defined by multiplying the diffractive
structure function by an appropriate power of $\xpom$ will not be
independent of $\xpom$. Hence, for such data we cannot define a
``pomeron structure function'', $\ftwopom$. In order to have a
meaningful comparison with $\ftwopom$ data, therefore, it is necessary
to consider the diffractive structure function at fixed $\xpom$. In
the fits shown in
Figs.\,\ref{fig:erdlf2d2108},\,\ref{fig:cminusscalarf2d2108} and
\ref{fig:2gluef2d2108}, data are plotted for which $\xpom\approx
0.0065$. The model curves are calculated at the corresponding $\xpom$
values, and, where there are no data in $\qsq$ and $\beta$ with
appropriate $\xpom$, the theory points are calculated using the value
$\xpom=0.00649$, which is the most common value of $\xpom$ of the
plotted data points.

For comparison, we show in Figs.\,\ref{fig:cminusscalar108multigraph}
and \ref{fig:cminusscalarf2d2108} the equivalent fits for the case of
the colour-singlet part of single-gluon exchange (solid line) and the
pomeron-quark coupling contribution to the scalar pomeron model
(dotted line).

\begin{figure}[htp]
\begin{center} \begin{picture}(1.5,1.5)(0,0)
\SetScale{72}
\SetOffset(-200,-440)
\SetWidth{.001}
\LogAxis(0,0)(1.5,0)(3,.040,1,.0100)
\LogAxis(0,0)(0,1.5)(4,-.040,1,.0100)
 \SetColor{BlueViolet}
 \SetColor{Emerald}
\Curve{(.7386,1.13)(.9515,1.02)(1.301,.86)}
 \SetColor{BrickRed}
 \SetColor{RawSienna}
\DashCurve{(.7386,1.22)(.9515,0.95)(1.301,.53)}{.03}
 \SetColor{Black}
\Text(37,-7)[]{\scriptsize $10^{-3}$}
\Text(74,-7)[]{\scriptsize $10^{-2}$}
\Text(-11,27)[]{\scriptsize $10^{-1}$}
\Text(-11,54)[]{\scriptsize $10^{0}$}
\Text(-11,80)[]{\scriptsize $10^{1}$}

\Text(7,20)[l]{\scriptsize $Q^2=8.5\,{\rm GeV}^2$}
\Text(7,11)[l]{\scriptsize $\beta=0.65$}

\SetOffset(-200,-332)
\SetWidth{.001}
\LogAxis(0,0)(1.5,0)(3,.040,1,.0100)
\LogAxis(0,0)(0,1.5)(4,-.040,1,.0100)
\SetColor{BlueViolet}
\SetColor{Emerald}
\Curve{(.7386,1.02)(.9515,.89)(1.301,0.50)}
\SetColor{BrickRed}
\SetColor{RawSienna}
\DashCurve{(.7386,1.21)(.9515,.92)(1.301,.27)}{.03}
\SetColor{Black}
\Text(-11,27)[]{\scriptsize $10^{-1}$}
\Text(-11,54)[]{\scriptsize $10^{0}$}
\Text(-11,80)[]{\scriptsize $10^{1}$}

\Text(7,20)[l]{\scriptsize $Q^2=8.5\,{\rm GeV}^2$}
\Text(7,11)[l]{\scriptsize $\beta=0.375$}

\SetOffset(-200,-224)
\SetWidth{.001}
\LogAxis(0,0)(1.5,0)(3,.040,1,.0100)
\LogAxis(0,0)(0,1.5)(4,-.040,1,.0100)
\SetColor{BlueViolet}
\SetColor{Emerald}
\Curve{(.7386,0.94)(.9515,.79)(1.301,.40)}
\SetColor{BrickRed}
\SetColor{RawSienna}
\DashCurve{(.7386,1.14)(.9515,.85)(1.301,.21)}{.03}
\SetColor{Black}
\Text(-11,27)[]{\scriptsize $10^{-1}$}
\Text(-11,54)[]{\scriptsize $10^{0}$}
\Text(-11,80)[]{\scriptsize $10^{1}$}

\Text(7,20)[l]{\scriptsize $Q^2=8.5\,{\rm GeV}^2$}
\Text(7,11)[l]{\scriptsize $\beta=0.175$}

\SetOffset(-200,-116)
\SetWidth{.001}
\LogAxis(0,0)(1.5,0)(3,.040,1,.0100)
\LogAxis(0,1.5)(1.5,1.5)(4,-.040,1,.0100) %
\LogAxis(0,0)(0,1.5)(4,-.040,1,.0100)
\Vertex(.7811,.969){.014}
\Curve{(.7811,.707)(.7811,1.065)}
\Vertex(.9061,.995){.014}
\Curve{(.9061,.832)(.9061,1.074)}
\SetColor{BlueViolet}
\SetColor{Emerald}
\Curve{(.6505,0.88)(.7811,0.81)(.9061,0.73)(.9515,.70)(1.1505,.48)}
\SetColor{BrickRed}
\SetColor{RawSienna}
\DashCurve{(.6505,1.14)(.7811,0.98)(.9061,0.82)(.9515,.76)(1.1505,.42)}{.03}

\SetColor{Black}

\Text(-11,27)[]{\scriptsize $10^{-1}$}
\Text(-11,54)[]{\scriptsize $10^{0}$}
\Text(-11,80)[]{\scriptsize $10^{1}$}

\Text(7,20)[l]{\scriptsize $Q^2=8.5\,{\rm GeV}^2$}
\Text(7,11)[l]{\scriptsize $\beta=0.065$}

\rText(-27,82)[][l]{\small $\fd$}

\SetOffset(-92,-440)
\SetWidth{.001}
\LogAxis(0,0)(1.5,0)(3,.040,1,.0100)
\LogAxis(0,0)(0,1.5)(4,-.040,1,.0100)
\Vertex(.9061,.990){.014}
\Curve{(.9061,.854)(.9061,1.063)}
\SetColor{BlueViolet}
\SetColor{Emerald}
\Curve{(.7386,1.14)(.9061,1.06)(1.301,.87)(1.4515,0.81)}
\SetColor{BrickRed}
\SetColor{RawSienna}
\DashCurve{(.7386,1.22)(.9061,1.02)(1.301,.54)(1.4515,0.37)}{.03}
\SetColor{Black}
\Text(37,-7)[]{\scriptsize $10^{-3}$}
\Text(74,-7)[]{\scriptsize $10^{-2}$}
\Text(7,20)[l]{\scriptsize $Q^2=12\,{\rm GeV}^2$}
\Text(7,11)[l]{\scriptsize $\beta=0.65$}

\SetOffset(-92,-332)
\SetWidth{.001}
\LogAxis(0,0)(1.5,0)(3,.040,1,.0100)
\LogAxis(0,0)(0,1.5)(4,-.040,1,.0100)
\Vertex(.7757,1.125){.014}
\Curve{(.7757,1.065)(.7757,1.168)}
\Vertex(.9004,1.032){.014}
\Curve{(.9004,.974)(.9004,1.075)}
\Vertex(1.0256,.955){.014}
\Curve{(1.0256,.865)(1.0256,1.013)}
\SetColor{BlueViolet}
\SetColor{Emerald}
\Curve{(.6505,1.08)(.7757,1.02)(.9004,0.95)(1.0256,.86)(1.301,0.56)}
\SetColor{BrickRed}
\SetColor{RawSienna}
\DashCurve{(.6505,1.33)(.7757,1.18)(.9004,1.01)(1.0256,.84)(1.301,0.33)}{.03}
\SetColor{Black}
\Text(7,20)[l]{\scriptsize $Q^2=12\,{\rm GeV}^2$}
\Text(7,11)[l]{\scriptsize $\beta=0.375$}

\SetOffset(-92,-224)
\SetWidth{.001}
\LogAxis(0,0)(1.5,0)(3,.040,1,.0100)
\LogAxis(0,0)(0,1.5)(4,-.040,1,.0100)
\Vertex(.8162,1.076){.014}
\Curve{(.8162,.980)(.8162,1.136)}
\Vertex(.9410,.943){.014}
\Curve{(.9410,.782)(.9410,1.023)}
\Vertex(1.0660,.886){.014}
\Curve{(1.0660,.809)(1.0660,.938)}
\Vertex(1.1910,.716){.014}
\Curve{(1.1910,.405)(1.1910,.816)}
\SetColor{BlueViolet}
\SetColor{Emerald}
\Curve{(.6505,1.00)(.8162,0.91)(.9410,.83)(1.066,.71)(1.191,.54)(1.239,.41)}
\SetColor{BrickRed}
\SetColor{RawSienna}
\DashCurve{(.6505,1.26)(.8162,1.06)(.9410,.89)(1.066,.70)(1.191,.43)(1.239,.27)}{.03}
\SetColor{Black}

\Text(7,20)[l]{\scriptsize $Q^2=12\,{\rm GeV}^2$}
\Text(7,11)[l]{\scriptsize $\beta=0.175$}

\SetOffset(-92,-116)
\SetWidth{.001}
\LogAxis(0,0)(1.5,0)(3,.040,1,.0100)
\LogAxis(0,1.5)(1.5,1.5)(4,-.040,1,.0100) %
\LogAxis(0,0)(0,1.5)(4,-.040,1,.0100)
\Vertex(.9061,.986){.014}
\Curve{(.9061,.884)(.9061,1.048)}
\Vertex(1.0311,.943){.014}
\Curve{(1.0311,.777)(1.0311,1.024)}
\Vertex(1.1561,.690){.014}
\Curve{(1.1561,.518)(1.1561,.771)}
\Vertex(1.2810,.761){.014}
\Curve{(1.2810,.644)(1.2810,.829)}
\SetColor{BlueViolet}
\SetColor{Emerald}
\Curve{(.6990,0.87)(.9061,0.75)(1.0311,.66)(1.1561,.52)(1.2810,.22)}
\SetColor{BrickRed}
\SetColor{RawSienna}
\DashCurve{(.6990,1.09)(.9061,0.84)(1.0311,.66)(1.1561,.45)(1.2810,.07)}{.03}
\SetColor{Black}
\Text(7,20)[l]{\scriptsize $Q^2=12\,{\rm GeV}^2$}
\Text(7,11)[l]{\scriptsize $\beta=0.065$}

\SetOffset(16,-440)
\SetWidth{.001}
\LogAxis(0,0)(1.5,0)(3,.040,1,.0100)
\LogAxis(0,0)(0,1.5)(4,-.040,1,.0100)
\Vertex(.9061,1.057){.014}
\Curve{(.9061,.977)(.9061,1.110)}
\Vertex(1.0311,.926){.014}
\Curve{(1.0311,.819)(1.0311,.989)}
\SetColor{BlueViolet}
\SetColor{Emerald}
\Curve{(.6505,1.20)(.7386,1.16)(.9061,1.09)(1.031,1.03)}
\SetColor{BrickRed}
\SetColor{RawSienna}
\DashCurve{(.6505,1.35)(.7386,1.25)(.9061,1.05)(1.031,.90)}{.03}
\SetColor{Black}
\Text(37,-7)[]{\scriptsize $10^{-3}$}
\Text(74,-7)[]{\scriptsize $10^{-2}$}
\Text(7,20)[l]{\scriptsize $Q^2=25\,{\rm GeV}^2$}
\Text(7,11)[l]{\scriptsize $\beta=0.65$}

\SetOffset(16,-332)
\SetWidth{.001}
\LogAxis(0,0)(1.5,0)(3,.040,1,.0100)
\LogAxis(0,0)(0,1.5)(4,-.040,1,.0100)
\Vertex(.7757,1.126){.014}
\Curve{(.7757,1.070)(.7757,1.168)}
\Vertex(.9004,1.000 ){.014}
\Curve{(.9004,0.938)(.9004,1.045)}
\Vertex(1.0256,0.932){.014}
\Curve{(1.0256,.868 )(1.0256,.978 )}
\Vertex(1.1505,.816 ){.014}
\Curve{(1.1505,.710 )(1.1505,.880 )}
\SetColor{BlueViolet}
\SetColor{Emerald}
\Curve{(.6505,1.10)(.7757,1.05)(.9004,0.98)(1.0256,.91)(1.1505,.82)(1.301,.67)}
\SetColor{BrickRed}
\SetColor{RawSienna}
\DashCurve{(.6505,1.35)(.7757,1.20)(.9004,1.05)(1.0256,.88)(1.1505,.70)(1.301,.44)}{.03}
\SetColor{Black}
\Text(7,20)[l]{\scriptsize $Q^2=25\,{\rm GeV}^2$}
\Text(7,11)[l]{\scriptsize $\beta=0.375$}

\SetOffset(16,-224)
\SetWidth{.001}
\LogAxis(0,0)(1.5,0)(3,.040,1,.0100)
\LogAxis(0,0)(0,1.5)(4,-.040,1,.0100)
\Vertex(.8162,1.148){.014}
\Curve{(.8162,1.081)(.8162,1.196)}
\Vertex(.9410,.944){.014}
\Curve{(.9410,.863)(.9410,.998)}
\Vertex(1.0660,.880){.014}
\Curve{(1.0660,.752)(1.0660,.951)}
\Vertex(1.1910,.837){.014}
\Curve{(1.1910,.766)(1.1910,.887)}
\Vertex(1.3160,.677){.014}
\Curve{(1.3160,.511)(1.3160,.758)}
\SetColor{BlueViolet}
\SetColor{Emerald}
\Curve{(.6505,1.01)(0.8162,0.94)(0.9410,.87)(1.0660,.78)(1.191,.65)(1.3160,.40)}
\SetColor{BrickRed}
\SetColor{RawSienna}
\DashCurve{(.6505,1.29)(0.8162,1.09)(0.9410,.93)(1.0660,.75)(1.191,.54)(1.3160,.21)}{.03}
\SetColor{Black}
\Text(7,20)[l]{\scriptsize $Q^2=25\,{\rm GeV}^2$}
\Text(7,11)[l]{\scriptsize $\beta=0.175$}

\SetOffset(16,-116)
\SetWidth{.001}
\LogAxis(0,0)(1.5,0)(3,.040,1,.0100)
\LogAxis(0,1.5)(1.5,1.5)(4,-.040,1,.0100) %
\LogAxis(0,0)(0,1.5)(4,-.040,1,.0100)
\Vertex(1.0311,.878){.014}
\Curve{(1.0311,.774)(1.0311,.940)}
\Vertex(1.1561,.722){.014}
\Curve{(1.1561,.605)(1.1561,.789)}
\Vertex(1.2810,.669){.014}
\Curve{(1.2810,.524)(1.2810,.745)}
\Vertex(1.4061,0){.014}
\Curve{(1.4061,0)(1.4061,.418)}
\SetColor{BlueViolet}
\SetColor{Emerald}
\Curve{(.8266,0.83)(1.0311,0.71)(1.1561,.60)(1.2810,.44)(1.3495,0.26)(1.4061,.06)}
\SetColor{BrickRed}
\SetColor{RawSienna}
\DashCurve{(.8266,0.97)(1.0311,0.71)(1.1561,.52)(1.2810,.28)(1.3495,0.06)}{.03}
\SetColor{Black}
\Text(7,20)[l]{\scriptsize $Q^2=25\,{\rm GeV}^2$}
\Text(7,11)[l]{\scriptsize $\beta=0.065$}

\SetOffset(124,-440)
\SetWidth{.001}
\LogAxis(0,0)(1.5,0)(3,.040,1,.0100)
\LogAxis(0,0)(0,1.5)(4,-.040,1,.0100)
\LogAxis(1.5,0)(1.5,1.5)(3,.040,0,.0100) %
\Vertex(.9061,1.118){.014}
\Curve{(.9061,1.012)(.9061,1.181)}
\Vertex(1.0311,.905){.014}
\Curve{(1.0311,.716)(1.0311,.990)}
\Vertex(1.1561,.815){.014}
\Curve{(1.1561,.647)(1.1561,.896)}
\SetColor{BlueViolet}
\SetColor{Emerald}
\Curve{(.6505,1.21)(.9061,1.11)(1.0311,1.05)(1.1561,.99)(1.301,.92)}
\SetColor{BrickRed}
\SetColor{RawSienna}
\DashCurve{(.6505,1.37)(.9061,1.08)(1.0311,.92)(1.1561,.77)(1.301,.59)}{.03}
\SetColor{Black}
\Text(37,-7)[]{\scriptsize $10^{-3}$}
\Text(74,-7)[]{\scriptsize $10^{-2}$}
\Text(7,20)[l]{\scriptsize $Q^2=50\,{\rm GeV}^2$}
\Text(7,11)[l]{\scriptsize $\beta=0.65$}

\Text(88,-21)[]{\small $x_{I\!\!P}$}

\SetOffset(124,-332)
\SetWidth{.001}
\LogAxis(0,0)(1.5,0)(3,.040,1,.0100)
\LogAxis(0,0)(0,1.5)(4,-.040,1,.0100)
\LogAxis(1.5,0)(1.5,1.5)(3,.040,0,.0100) %
\Vertex(.7757,1.173){.014}
\Curve{(.7757,1.064)(.7757,1.237)}
\Vertex(.9004,1.127){.014}
\Curve{(.9004,1.038)(.9004,1.184)}
\Vertex(1.0256,.996){.014}
\Curve{(1.0256,.921)(1.0256,1.048)}
\Vertex(1.1505,.866){.014}
\Curve{(1.1505,.761)(1.1505,.929)}
\Vertex(1.2755,.802){.014}
\Curve{(1.2755,.496)(1.2755,.902)}
\SetColor{BlueViolet}
\SetColor{Emerald}
\Curve{(.6505,1.11)(.7757,1.07)(0.9004,1.01)(1.0256,.95)(1.1505,.86)(1.2755,.76)(1.301,.74)(1.4515,.55)}
\SetColor{BrickRed}
\SetColor{RawSienna}
\DashCurve{(.6505,1.37)(.7757,1.23)(0.9004,1.08)(1.0256,.92)(1.1505,.74)(1.2755,.55)(1.301,.51)(1.4515,.20)}{.03}
\SetColor{Black}

\Text(7,20)[l]{\scriptsize $Q^2=50\,{\rm GeV}^2$}
\Text(7,11)[l]{\scriptsize $\beta=0.375$}

\SetOffset(124,-224)
\SetWidth{.001}
\LogAxis(0,0)(1.5,0)(3,.040,1,.0100)
\LogAxis(0,0)(0,1.5)(4,-.040,1,.0100)
\LogAxis(1.5,0)(1.5,1.5)(3,.040,0,.0100) %
\Vertex(.9410,1.035){.014}
\Curve{(.9410,.899)(.9410,1.108)}
\Vertex(1.0660,.906){.014}
\Curve{(1.0660,.729)(1.0660,.989)}
\Vertex(1.1910,.812){.014}
\Curve{(1.1910,.664)(1.1910,.888)}
\Vertex(1.3160,.634){.014}
\Curve{(1.3160,.317)(1.3160,.735)}
\Vertex(1.4410,.317){.014}
\Curve{(1.4410,0)(1.4410,.627)}
\SetColor{BlueViolet}
\SetColor{Emerald}
\Curve{(.8010,0.97)(0.9410,0.91)(1.0660,.83)(1.1910,.72)(1.3160,.57)(1.3891,.40)}
\SetColor{BrickRed}
\SetColor{RawSienna}
\DashCurve{(.8010,1.13)(0.9410,0.96)(1.0660,.80)(1.1910,.61)(1.3160,.37)(1.3891,.15)}{.03}
\SetColor{Black}
\Text(7,20)[l]{\scriptsize $Q^2=50\,{\rm GeV}^2$}
\Text(7,11)[l]{\scriptsize $\beta=0.175$}

\SetOffset(124,-116)
\SetWidth{.001}
\LogAxis(0,0)(1.5,0)(4,-.040,1,.0100)
\LogAxis(0,0)(0,1.5)(3,.040,1,.0100)

\end{picture}
\vspace*{6.3in}
\caption[Structure function $\fd$ for the single gluon exchange and
scalar pomeron models.]  {{\it The single-gluon exchange and
scalar-pomeron models are fitted to virtuality-cut
data~\cite{Phillips:1995jpp} for $\fd$, with parameters as in
Fig.\,\ref{fig:erdl108multigraph}. The solid line is the prediction
from the single-gluon exchange model, and the dashed line is the
scalar pomeron model.}}
\label{fig:cminusscalar108multigraph}
\end{center}
\end{figure}

\begin{figure}[htp]
\begin{center} \begin{picture}(2.5,1.25)(0,-20)
\SetScale{72}  
\SetOffset(-200,-440)
\SetWidth{.001}
\LinAxis(0,0)(0,1.25)(5,2,-.040,0,.0100)
\LogAxis(0,0)(2.5,0)(1.5,.040,5,.0100)
\SetColor{Black}
\SetScale{1}
\Vertex(45.4,34.02){1}
\Curve{(45.4,14.67)(45.4,53.37)}
\Vertex(83.4,51.12){1}
\Curve{(83.4,31.14)(83.4,70.83)}
\Vertex(119,74.43){1}
\Curve{(119,38.88)(119,90)}
\SetColor{Emerald}
\Curve{(27.5,54.73)(45.4,60.20)(83.4,71.53)(119,83.39)}
\SetColor{BrickRed}
\SetColor{RawSienna}
\DashCurve{(27.5,44.15)(45.4,48.41)(83.4,57.20)(119,66.44)}{3}
\SetScale{72}
\SetColor{Black}
\Text(37,-7)[]{\scriptsize $10$}
\Text(155,-7)[]{\scriptsize $10^{2}$}
\Text(165,-17)[]{\scriptsize $Q^2\,/{\rm GeV}^2$}
\Text(155,-7)[]{\scriptsize $10^{2}$}

\Text(-8,0)[]{\scriptsize $0$}
\Text(-8,18)[]{\scriptsize $2$}
\Text(-8,36)[]{\scriptsize $4$}
\Text(-8,54)[]{\scriptsize $6$}
\Text(-8,72)[]{\scriptsize $8$}
\Text(90,-29)[]{\scriptsize $(a)$}

\Text(12,74)[l]{\scriptsize $\beta=0.65$}
\LinAxis(2.5,0)(2.5,1.25)(5,2,.040,0,.0100)

\SetOffset(-200,-350)
\SetWidth{.001}
\LinAxis(0,0)(0,1.25)(5,2,-.040,0,.0100)
\LogAxis(0,0)(2.5,0)(1.5,.040,5,.0100)
\SetScale{1}
\Vertex(45.4,42.39){1}
\Curve{(45.4,29.7)(45.4,55.08)}
\Vertex(83.4,34.94){1}
\Curve{(83.4,23.85)(83.4,45.99)}
\Vertex(119,76.05){1}
\Curve{(119,44.01)(119,90)}
\SetColor{Emerald}
\Curve{(27.5,26.37)(45.4,30.30)(83.4,37.78)(119,45.09)}
\SetColor{BrickRed}
\SetColor{RawSienna}
\DashCurve{(27.5,38.47)(45.4,45.26)(83.4,56.24)(119,67.26)}{3}
\SetColor{Black}
\SetScale{72}

\Text(-8,0)[]{\scriptsize $0$}
\Text(-8,18)[]{\scriptsize $2$}
\Text(-8,36)[]{\scriptsize $4$}
\Text(-8,54)[]{\scriptsize $6$}
\Text(-8,72)[]{\scriptsize $8$}

\Text(12,74)[l]{\scriptsize $\beta=0.375$}

\LinAxis(2.5,0)(2.5,1.25)(5,2,.040,0,.0100)

\SetOffset(-200,-260)
\SetWidth{.001}
\LinAxis(0,0)(0,1.25)(5,2,-.040,0,.0100)
\LogAxis(0,0)(2.5,0)(1.5,.040,5,.0100)
\SetScale{1}
\Vertex(45.4,31.95){1}
\Curve{(45.4,11.88)(45.4,52.02)}
\Vertex(83.4,32.04){1}
\Curve{(83.4,19.44)(83.4,44.64)}
\Vertex(119,55.98){1}
\Curve{(119,24.3)(119,87.66)}
\SetColor{Emerald}
\Curve{(27.5,14.65)(45.4,14.38)(83.4,19.05)(119,23.46)}
\SetColor{BrickRed}
\SetColor{RawSienna}
\DashCurve{(27.5,25.61)(45.4,21.40)(83.4,27.41)(119,33.24)}{3}
\SetColor{Black}
\SetScale{72}
 
\Text(-8,0)[]{\scriptsize $0$}
\Text(-8,18)[]{\scriptsize $2$}
\Text(-8,36)[]{\scriptsize $4$}
\Text(-8,54)[]{\scriptsize $6$}
\Text(-8,72)[]{\scriptsize $8$}
 
\Text(12,74)[l]{\scriptsize $\beta=0.175$}
\LinAxis(2.5,0)(2.5,1.25)(5,2,.040,0,.0100)
 
\SetOffset(-200,-170)
\SetWidth{.001}
\LinAxis(0,0)(0,1.25)(5,2,-.040,0,.0100)
\LogAxis(0,0)(2.5,0)(1.5,.040,5,.0100)
\SetScale{1}
\Vertex(106,80){1}
\SetColor{Emerald}
\Curve{(102,72)(110,72)}
\SetColor{BrickRed}
\SetColor{RawSienna}
\DashCurve{(102,64)(110,64)}{3}
\SetColor{Black}
\Vertex(27.5,34.92){1}
\Curve{(27.5,12.78)(27.5,57.06)}
\Vertex(45.4,33.21){1}
\Curve{(45.4,17.73)(45.4,48.69)}
\SetColor{Emerald}
\Curve{(27.5,8.140)(45.4,9.260)}
\SetColor{BrickRed}
\SetColor{RawSienna}
\DashCurve{(27.5,13.98)(45.4,15.53)}{3}
\SetColor{Black}
\SetScale{72}
 
\Text(-8,0)[]{\scriptsize $0$}
\Text(-8,18)[]{\scriptsize $2$}
\Text(-8,36)[]{\scriptsize $4$}
\Text(-8,54)[]{\scriptsize $6$}
\Text(-8,72)[]{\scriptsize $8$}
\rText(-21,72)[][l]{\scriptsize $F_{2}^{\pom}$}
 
\Text(121,80)[l]{\scriptsize\em H1 Data}
\Text(121,72)[l]{\scriptsize\em Single gluon}
\Text(121,64)[l]{\scriptsize\em Scalar pomeron}
 
\Text(12,74)[l]{\scriptsize $\beta=0.065$}
 
\LinAxis(2.5,0)(2.5,1.25)(5,2,.040,0,.0100)
\LogAxis(0,1.25)(2.5,1.25)(1.5,-.040,5,.0100)
 
\SetOffset(26,-440)
\SetWidth{.001}
\LinAxis(0,0)(0,1.25)(5,2,-.040,0,.0100)
\LinAxis(0,0)(2.5,0)(5,2,.040,0,.0100)
\SetScale{1}
\Vertex(31.5,55.98){1}
\Curve{(31.5,24.3)(31.5,87.66)}
\Vertex(67.5,76.05){1}
\Curve{(67.5,44.01)(67.5,90)}
\Vertex(117,74.43){1}
\Curve{(117,38.88)(117,90)}
\SetColor{Emerald}
\Curve{(31.5,23.46)(67.5,45.09)(117,83.39)}
\SetColor{BrickRed}
\SetColor{RawSienna}
\DashCurve{(31.5,33.24)(67.5,67.26)(117,66.44)}{3}
\SetScale{72}
\SetColor{Black}
\Text(0,-7)[]{\scriptsize $0$}
\Text(36,-7)[]{\scriptsize $0.2$}
\Text(72,-7)[]{\scriptsize $0.4$}
\Text(108,-7)[]{\scriptsize $0.6$}
\Text(144,-7)[]{\scriptsize $0.8$}
\Text(180,-7)[]{\scriptsize $1$}
\Text(165,-17)[]{\scriptsize $\beta$}
\Text(90,-29)[]{\scriptsize $(b)$}
 
\Text(-8,0)[]{\scriptsize $0$}
\Text(-8,18)[]{\scriptsize $2$}
\Text(-8,36)[]{\scriptsize $4$}
\Text(-8,54)[]{\scriptsize $6$}
\Text(-8,72)[]{\scriptsize $8$}
 
\Text(12,74)[l]{\scriptsize $Q^2=50\,{\rm GeV}^2$}
\LinAxis(2.5,0)(2.5,1.25)(5,2,.040,0,.0100)

\SetOffset(26,-350)
\SetWidth{.001}
\LinAxis(0,0)(0,1.25)(5,2,-.040,0,.0100)
\LinAxis(0,0)(2.5,0)(5,2,.040,0,.0100)
\SetScale{1}
\Vertex(31.5,32.04){1}
\Curve{(31.5,19.44)(31.5,44.64)}
\Vertex(67.5,34.92){1}
\Curve{(67.5,23.85)(67.5,45.99)}
\Vertex(117,51.12){1}
\Curve{(117,31.41)(117,70.83)}
\SetColor{Emerald}
\Curve{(31.5,19.05)(67.5,37.78)(117,71.53)}
\SetColor{BrickRed}
\SetColor{RawSienna}
\DashCurve{(31.5,27.41)(67.5,56.24)(117,57.20)}{3}
\SetColor{Black}
\SetScale{72}
 
\Text(-8,0)[]{\scriptsize $0$}
\Text(-8,18)[]{\scriptsize $2$}
\Text(-8,36)[]{\scriptsize $4$}
\Text(-8,54)[]{\scriptsize $6$}
\Text(-8,72)[]{\scriptsize $8$}
 
\Text(12,74)[l]{\scriptsize $Q^2=25\,{\rm GeV}^2$}
\LinAxis(2.5,0)(2.5,1.25)(5,2,.040,0,.0100)

\SetOffset(26,-260)
\SetWidth{.001}
\LinAxis(0,0)(0,1.25)(5,2,-.040,0,.0100)
\LinAxis(0,0)(2.5,0)(5,2,.040,0,.0100)
\SetScale{1}
\Vertex(11.7,33.21){1}
\Curve{(11.7,17.73)(11.7,48.69)}
\Vertex(31.5,31.95){1}
\Curve{(31.5,11.88)(31.5,52.02)}
\Vertex(67.5,42.39){1}
\Curve{(67.5,29.7)(67.5,55.08)}
\Vertex(117,34.02){1}
\Curve{(117,14.67)(117,53.37)}
\SetColor{Emerald}
\Curve{(11.7,09.26)(31.5,14.38)(67.5,30.30)(117,60.20)}
\SetColor{BrickRed}
\SetColor{RawSienna}
\DashCurve{(11.7,15.53)(31.5,21.40)(67.5,45.26)(117,48.41)}{3}
\SetColor{Black}
\SetScale{72}
 
\Text(-8,0)[]{\scriptsize $0$}
\Text(-8,18)[]{\scriptsize $2$}
\Text(-8,36)[]{\scriptsize $4$}
\Text(-8,54)[]{\scriptsize $6$}
\Text(-8,72)[]{\scriptsize $8$}
 
\Text(12,74)[l]{\scriptsize $Q^2=12\,{\rm GeV}^2$}
\LinAxis(2.5,0)(2.5,1.25)(5,2,.040,0,.0100)

\SetOffset(26,-170)
\SetWidth{.001}
\LinAxis(0,0)(0,1.25)(5,2,-.040,0,.0100)
\LinAxis(0,0)(2.5,0)(5,2,.040,0,.0100)
\SetScale{1}
 
\Vertex(106,80){1}
\SetColor{Emerald}
\Curve{(102,72)(110,72)}
\SetColor{BrickRed}
\SetColor{RawSienna}
\DashCurve{(102,64)(110,64)}{3}
\SetColor{Black}
\Vertex(11.7,34.92){1}
\Curve{(11.7,12.78)(11.7,57.06)}
\SetColor{Emerald}
\Curve{(11.7,08.14)(31.5,14.65)(67.5,26.37)(117,54.73)}
\SetColor{BrickRed}
\SetColor{RawSienna}
\DashCurve{(11.7,13.98)(31.5,25.61)(67.5,38.47)(117,44.15)}{3}
\SetColor{Black}
\SetScale{72}
 
\Text(-8,0)[]{\scriptsize $0$}
\Text(-8,18)[]{\scriptsize $2$}
\Text(-8,36)[]{\scriptsize $4$}
\Text(-8,54)[]{\scriptsize $6$}
\Text(-8,72)[]{\scriptsize $8$}
\rText(-21,72)[][l]{\scriptsize $F_{2}^{\pom}$}
 
\Text(121,80)[l]{\scriptsize\em H1 Data}
\Text(121,72)[l]{\scriptsize\em Single gluon}
\Text(121,64)[l]{\scriptsize\em Scalar pomeron}
 
\Text(12,74)[l]{\scriptsize $Q^2=8.5\,{\rm GeV}^2$}
\LinAxis(2.5,0)(2.5,1.25)(5,2,.040,0,.0100)
\LinAxis(0,1.25)(2.5,1.25)(5,2,-.040,0,.0100)

\end{picture}
\vspace*{6.3in}
\caption[Structure function $\ftwopom$ for the single gluon exchange
and scalar pomeron models.]  {{\it The single-gluon exchange and
scalar-pomeron models are fitted to virtuality cut
data~\cite{Phillips:1995jpp} for $\ftwopom$, with parameters as in
Fig.\,\ref{fig:erdl108multigraph}. The solid line is the prediction
from the single-gluon exchange model, and the dashed line is the
scalar pomeron model.} \label{fig:cminusscalarf2d2108}}
\end{center}
\end{figure}

In Figs.\,\ref{fig:2glue108multigraph} and\,\ref{fig:2gluef2d2108} we
show the fits using the two-gluon exchange model of
Diehl\,\cite{Diehl:1995wz}. The most striking feature here is the
failure of the model to describe the small-$\beta$ dependence of the
pseudo-rapidity cut data. The reason for this may be seen in the
analytic form presented by equation~(30)
of\,\cite{Diehl:1995wz}, which shows that, for large $\ptsqmin$,
there is a $\beta^3$ dependence, in disagreement
with the rather flat dependence found experimentally. Only if the
$\ptsqmin$ cut is removed does the two-gluon term generate an
approximately linear $\beta$ dependence.

\begin{figure}[htp]
\begin{center} \begin{picture}(1.5,1.5)(0,0)
\SetScale{72}
\SetOffset(-200,-440)
\SetWidth{.001}
\LogAxis(0,0)(1.5,0)(3,.040,1,.0100)
\LogAxis(0,0)(0,1.5)(4,-.040,1,.0100)
 \SetColor{BlueViolet}
 \SetColor{Magenta}
\Curve{(.7386,1.49)(.9515,1.09)(1.301,.59)}
 \SetColor{BrickRed}
 \SetColor{Black}
\Text(37,-7)[]{\scriptsize $10^{-3}$}
\Text(74,-7)[]{\scriptsize $10^{-2}$}
\Text(-11,27)[]{\scriptsize $10^{-1}$}
\Text(-11,54)[]{\scriptsize $10^{0}$}
\Text(-11,80)[]{\scriptsize $10^{1}$}

\Text(7,20)[l]{\scriptsize $Q^2=8.5\,{\rm GeV}^2$}
\Text(7,11)[l]{\scriptsize $\beta=0.65$}

\SetOffset(-200,-332)
\SetWidth{.001}
\LogAxis(0,0)(1.5,0)(3,.040,1,.0100)
\LogAxis(0,0)(0,1.5)(4,-.040,1,.0100)
\SetColor{BlueViolet}
\SetColor{Magenta}
\Curve{(.7386,1.18)(.9515,.75)(1.301,0.02)}
\SetColor{BrickRed}
\SetColor{Black}
\Text(-11,27)[]{\scriptsize $10^{-1}$}
\Text(-11,54)[]{\scriptsize $10^{0}$}
\Text(-11,80)[]{\scriptsize $10^{1}$}

\Text(7,20)[l]{\scriptsize $Q^2=8.5\,{\rm GeV}^2$}
\Text(7,11)[l]{\scriptsize $\beta=0.375$}

\SetOffset(-200,-224)
\SetWidth{.001}
\LogAxis(0,0)(1.5,0)(3,.040,1,.0100)
\LogAxis(0,0)(0,1.5)(4,-.040,1,.0100)
\SetColor{BlueViolet}
\SetColor{Magenta}
\Curve{(.7386,0.81)(.9515,.40)(1.15,0)}
\SetColor{BrickRed}
\SetColor{Black}
\Text(-11,27)[]{\scriptsize $10^{-1}$}
\Text(-11,54)[]{\scriptsize $10^{0}$}
\Text(-11,80)[]{\scriptsize $10^{1}$}

\Text(7,20)[l]{\scriptsize $Q^2=8.5\,{\rm GeV}^2$}
\Text(7,11)[l]{\scriptsize $\beta=0.175$}

\SetOffset(-200,-116)
\SetWidth{.001}
\LogAxis(0,0)(1.5,0)(3,.040,1,.0100)
\LogAxis(0,1.5)(1.5,1.5)(4,-.040,1,.0100) %
\LogAxis(0,0)(0,1.5)(4,-.040,1,.0100)
\Vertex(.7811,.969){.014}
\Curve{(.7811,.707)(.7811,1.065)}
\Vertex(.9061,.995){.014}
\Curve{(.9061,.832)(.9061,1.074)}
\SetColor{BlueViolet}
\SetColor{Magenta}
\Curve{(.6505,0.53)(.7811,0.28)(.9061,0.06)(.9515,-.01)}
\SetColor{BrickRed}
\SetColor{Black}

\Text(-11,27)[]{\scriptsize $10^{-1}$}
\Text(-11,54)[]{\scriptsize $10^{0}$}
\Text(-11,80)[]{\scriptsize $10^{1}$}

\Text(7,20)[l]{\scriptsize $Q^2=8.5\,{\rm GeV}^2$}
\Text(7,11)[l]{\scriptsize $\beta=0.065$}

\rText(-27,82)[][l]{\small $\fd$}

\SetOffset(-92,-440)
\SetWidth{.001}
\LogAxis(0,0)(1.5,0)(3,.040,1,.0100)
\LogAxis(0,0)(0,1.5)(4,-.040,1,.0100)
\Vertex(.9061,.990){.014}
\Curve{(.9061,.854)(.9061,1.063)}
\SetColor{BlueViolet}
\SetColor{Magenta}
\Curve{(.7386,1.46)(.9061,1.15)(1.301,.54)(1.4515,0.36)}
\SetColor{BrickRed}
\SetColor{Black}
\Text(37,-7)[]{\scriptsize $10^{-3}$}
\Text(74,-7)[]{\scriptsize $10^{-2}$}
\Text(7,20)[l]{\scriptsize $Q^2=12\,{\rm GeV}^2$}
\Text(7,11)[l]{\scriptsize $\beta=0.65$}

\SetOffset(-92,-332)
\SetWidth{.001}
\LogAxis(0,0)(1.5,0)(3,.040,1,.0100)
\LogAxis(0,0)(0,1.5)(4,-.040,1,.0100)
\Vertex(.7757,1.125){.014}
\Curve{(.7757,1.065)(.7757,1.168)}
\Vertex(.9004,1.032){.014}
\Curve{(.9004,.974)(.9004,1.075)}
\Vertex(1.0256,.955){.014}
\Curve{(1.0256,.865)(1.0256,1.013)}
\SetColor{BlueViolet}
\SetColor{Magenta}
\Curve{(.6505,1.36)(.7757,1.09)(.9004,0.83)(1.0256,.59)(1.301,0.03)}
\SetColor{BrickRed}
\SetColor{Black}
\Text(7,20)[l]{\scriptsize $Q^2=12\,{\rm GeV}^2$}
\Text(7,11)[l]{\scriptsize $\beta=0.375$}

\SetOffset(-92,-224)
\SetWidth{.001}
\LogAxis(0,0)(1.5,0)(3,.040,1,.0100)
\LogAxis(0,0)(0,1.5)(4,-.040,1,.0100)
\Vertex(.8162,1.076){.014}
\Curve{(.8162,.980)(.8162,1.136)}
\Vertex(.9410,.943){.014}
\Curve{(.9410,.782)(.9410,1.023)}
\Vertex(1.0660,.886){.014}
\Curve{(1.0660,.809)(1.0660,.938)}
\Vertex(1.1910,.716){.014}
\Curve{(1.1910,.405)(1.1910,.816)}
\SetColor{BlueViolet}
\SetColor{Magenta}
\Curve{(.6505,0.99)(.8162,0.64)(.9410,.40)(1.066,.17)}
\SetColor{BrickRed}
\SetColor{Black}

\Text(7,20)[l]{\scriptsize $Q^2=12\,{\rm GeV}^2$}
\Text(7,11)[l]{\scriptsize $\beta=0.175$}

\SetOffset(-92,-116)
\SetWidth{.001}
\LogAxis(0,0)(1.5,0)(3,.040,1,.0100)
\LogAxis(0,1.5)(1.5,1.5)(4,-.040,1,.0100) %
\LogAxis(0,0)(0,1.5)(4,-.040,1,.0100)
\Vertex(.9061,.986){.014}
\Curve{(.9061,.884)(.9061,1.048)}
\Vertex(1.0311,.943){.014}
\Curve{(1.0311,.777)(1.0311,1.024)}
\Vertex(1.1561,.690){.014}
\Curve{(1.1561,.518)(1.1561,.771)}
\Vertex(1.2810,.761){.014}
\Curve{(1.2810,.644)(1.2810,.829)}
\SetColor{BlueViolet}
\SetColor{Magenta}
\Curve{(.6990,0.42)(.9061,0.04)}
\SetColor{BrickRed}
\SetColor{Black}
\Text(7,20)[l]{\scriptsize $Q^2=12\,{\rm GeV}^2$}
\Text(7,11)[l]{\scriptsize $\beta=0.065$}

\SetOffset(16,-440)
\SetWidth{.001}
\LogAxis(0,0)(1.5,0)(3,.040,1,.0100)
\LogAxis(0,0)(0,1.5)(4,-.040,1,.0100)
\Vertex(.9061,1.057){.014}
\Curve{(.9061,.977)(.9061,1.110)}
\Vertex(1.0311,.926){.014}
\Curve{(1.0311,.819)(1.0311,.989)}
\SetColor{BlueViolet}
\SetColor{Magenta}
\Curve{(.716,1.5)(.7386,1.44)(.9061,1.10)(1.031,.87)}
\SetColor{BrickRed}
\SetColor{Black}
\Text(37,-7)[]{\scriptsize $10^{-3}$}
\Text(74,-7)[]{\scriptsize $10^{-2}$}
\Text(7,20)[l]{\scriptsize $Q^2=25\,{\rm GeV}^2$}
\Text(7,11)[l]{\scriptsize $\beta=0.65$}

\SetOffset(16,-332)
\SetWidth{.001}
\LogAxis(0,0)(1.5,0)(3,.040,1,.0100)
\LogAxis(0,0)(0,1.5)(4,-.040,1,.0100)
\Vertex(.7757,1.126){.014}
\Curve{(.7757,1.070)(.7757,1.168)}
\Vertex(.9004,1.000 ){.014}
\Curve{(.9004,0.938)(.9004,1.045)}
\Vertex(1.0256,0.932){.014}
\Curve{(1.0256,.868 )(1.0256,.978 )}
\Vertex(1.1505,.816 ){.014}
\Curve{(1.1505,.710 )(1.1505,.880 )}
\SetColor{BlueViolet}
\SetColor{Magenta}
\Curve{(.6505,1.35)(.7757,1.07)(.9004,0.80)(1.0256,.55)(1.1505,.31)(1.301,.02)}
\SetColor{BrickRed}
\SetColor{Black}
\Text(7,20)[l]{\scriptsize $Q^2=25\,{\rm GeV}^2$}
\Text(7,11)[l]{\scriptsize $\beta=0.375$}

\SetOffset(16,-224)
\SetWidth{.001}
\LogAxis(0,0)(1.5,0)(3,.040,1,.0100)
\LogAxis(0,0)(0,1.5)(4,-.040,1,.0100)
\Vertex(.8162,1.148){.014}
\Curve{(.8162,1.081)(.8162,1.196)}
\Vertex(.9410,.944){.014}
\Curve{(.9410,.863)(.9410,.998)}
\Vertex(1.0660,.880){.014}
\Curve{(1.0660,.752)(1.0660,.951)}
\Vertex(1.1910,.837){.014}
\Curve{(1.1910,.766)(1.1910,.887)}
\Vertex(1.3160,.677){.014}
\Curve{(1.3160,.511)(1.3160,.758)}
\SetColor{BlueViolet}
\SetColor{Magenta}
\Curve{(.6505,0.98)(0.8162,0.62)(0.9410,.37)(1.0660,.13)(1.139,0)}
\SetColor{BrickRed}
\SetColor{Black}
\Text(7,20)[l]{\scriptsize $Q^2=25\,{\rm GeV}^2$}
\Text(7,11)[l]{\scriptsize $\beta=0.175$}

\SetOffset(16,-116)
\SetWidth{.001}
\LogAxis(0,0)(1.5,0)(3,.040,1,.0100)
\LogAxis(0,1.5)(1.5,1.5)(4,-.040,1,.0100) %
\LogAxis(0,0)(0,1.5)(4,-.040,1,.0100)
\Vertex(1.0311,.878){.014}
\Curve{(1.0311,.774)(1.0311,.940)}
\Vertex(1.1561,.722){.014}
\Curve{(1.1561,.605)(1.1561,.789)}
\Vertex(1.2810,.669){.014}
\Curve{(1.2810,.524)(1.2810,.745)}
\Vertex(1.4061,0){.014}
\Curve{(1.4061,0)(1.4061,.418)}
\SetColor{BlueViolet}
\SetColor{Magenta}
\Curve{(.8266,0.13)(.889,0)}
\SetColor{BrickRed}
\SetColor{Black}
\Text(7,20)[l]{\scriptsize $Q^2=25\,{\rm GeV}^2$}
\Text(7,11)[l]{\scriptsize $\beta=0.065$}

\SetOffset(124,-440)
\SetWidth{.001}
\LogAxis(0,0)(1.5,0)(3,.040,1,.0100)
\LogAxis(0,0)(0,1.5)(4,-.040,1,.0100)
\LogAxis(1.5,0)(1.5,1.5)(3,.040,0,.0100) %
\Vertex(.9061,1.118){.014}
\Curve{(.9061,1.012)(.9061,1.181)}
\Vertex(1.0311,.905){.014}
\Curve{(1.0311,.716)(1.0311,.990)}
\Vertex(1.1561,.815){.014}
\Curve{(1.1561,.647)(1.1561,.896)}
\SetColor{BlueViolet}
\SetColor{Magenta}
\Curve{(.699,1.5)(.9061,1.07)(1.0311,.83)(1.1561,.61)(1.301,.39)}
\SetColor{BrickRed}
\SetColor{Black}
\Text(37,-7)[]{\scriptsize $10^{-3}$}
\Text(74,-7)[]{\scriptsize $10^{-2}$}
\Text(7,20)[l]{\scriptsize $Q^2=50\,{\rm GeV}^2$}
\Text(7,11)[l]{\scriptsize $\beta=0.65$}

\Text(88,-21)[]{\small $x_{I\!\!P}$}

\SetOffset(124,-332)
\SetWidth{.001}
\LogAxis(0,0)(1.5,0)(3,.040,1,.0100)
\LogAxis(0,0)(0,1.5)(4,-.040,1,.0100)
\LogAxis(1.5,0)(1.5,1.5)(3,.040,0,.0100) %
\Vertex(.7757,1.173){.014}
\Curve{(.7757,1.064)(.7757,1.237)}
\Vertex(.9004,1.127){.014}
\Curve{(.9004,1.038)(.9004,1.184)}
\Vertex(1.0256,.996){.014}
\Curve{(1.0256,.921)(1.0256,1.048)}
\Vertex(1.1505,.866){.014}
\Curve{(1.1505,.761)(1.1505,.929)}
\Vertex(1.2755,.802){.014}
\Curve{(1.2755,.496)(1.2755,.902)}
\SetColor{BlueViolet}
\SetColor{Magenta}
\Curve{(.6505,1.34)(.7757,1.06)(0.9004,0.78)(1.0256,.52)(1.1505,.28)(1.2755,.04)}
\SetColor{BrickRed}
\SetColor{Black}

\Text(7,20)[l]{\scriptsize $Q^2=50\,{\rm GeV}^2$}
\Text(7,11)[l]{\scriptsize $\beta=0.375$}

\SetOffset(124,-224)
\SetWidth{.001}
\LogAxis(0,0)(1.5,0)(3,.040,1,.0100)
\LogAxis(0,0)(0,1.5)(4,-.040,1,.0100)
\LogAxis(1.5,0)(1.5,1.5)(3,.040,0,.0100) %
\Vertex(.9410,1.035){.014}
\Curve{(.9410,.899)(.9410,1.108)}
\Vertex(1.0660,.906){.014}
\Curve{(1.0660,.729)(1.0660,.989)}
\Vertex(1.1910,.812){.014}
\Curve{(1.1910,.664)(1.1910,.888)}
\Vertex(1.3160,.634){.014}
\Curve{(1.3160,.317)(1.3160,.735)}
\Vertex(1.4410,.317){.014}
\Curve{(1.4410,0)(1.4410,.627)}
\SetColor{BlueViolet}
\SetColor{Magenta}
\Curve{(.8010,0.63)(0.9410,0.33)(1.0660,.09)(1.115,0)}
\SetColor{BrickRed}
\SetColor{Black}
\Text(7,20)[l]{\scriptsize $Q^2=50\,{\rm GeV}^2$}
\Text(7,11)[l]{\scriptsize $\beta=0.175$}

\SetOffset(124,-116)
\SetWidth{.001}
\LogAxis(0,0)(1.5,0)(4,-.040,1,.0100)
\LogAxis(0,0)(0,1.5)(3,.040,1,.0100)

\end{picture}
\vspace*{6.3in}
\caption[Structure function $\fd$ for the two gluon exchange model.]
{{\it The two-gluon exchange model is fitted to virtuality cut
data~\cite{Phillips:1995jpp} for $\fd$, with parameters as in
Fig.\,\ref{fig:erdl108multigraph}.}}
\label{fig:2glue108multigraph}
\end{center}
\end{figure}

\begin{figure}[htp]
\begin{center} \begin{picture}(2.5,1.25)(0,-20)
\SetScale{72}  
\SetOffset(-200,-440)
\SetWidth{.001}
\LinAxis(0,0)(0,1.25)(5,2,-.040,0,.0100)
\LogAxis(0,0)(2.5,0)(1.5,.040,5,.0100)
\SetColor{Black}
\SetScale{1}
\Vertex(45.4,34.02){1}
\Curve{(45.4,14.67)(45.4,53.37)}
\Vertex(83.4,51.12){1}
\Curve{(83.4,31.14)(83.4,70.83)}
\Vertex(119,74.43){1}
\Curve{(119,38.88)(119,90)}
\SetColor{Magenta}
\Curve{(62.1,90)(83.4,79.00)(119,65.13)}
\SetColor{BrickRed}
\SetScale{72}
\SetColor{Black}
\Text(37,-7)[]{\scriptsize $10$}
\Text(155,-7)[]{\scriptsize $10^{2}$}
\Text(165,-17)[]{\scriptsize $Q^2\,/{\rm GeV}^2$}
\Text(155,-7)[]{\scriptsize $10^{2}$}

\Text(-8,0)[]{\scriptsize $0$}
\Text(-8,18)[]{\scriptsize $2$}
\Text(-8,36)[]{\scriptsize $4$}
\Text(-8,54)[]{\scriptsize $6$}
\Text(-8,72)[]{\scriptsize $8$}
\Text(90,-29)[]{\scriptsize $(a)$}

\Text(12,74)[l]{\scriptsize $\beta=0.65$}
\LinAxis(2.5,0)(2.5,1.25)(5,2,.040,0,.0100)

\SetOffset(-200,-350)
\SetWidth{.001}
\LinAxis(0,0)(0,1.25)(5,2,-.040,0,.0100)
\LogAxis(0,0)(2.5,0)(1.5,.040,5,.0100)
\SetScale{1}
\Vertex(45.4,42.39){1}
\Curve{(45.4,29.7)(45.4,55.08)}
\Vertex(83.4,34.94){1}
\Curve{(83.4,23.85)(83.4,45.99)}
\Vertex(119,76.05){1}
\Curve{(119,44.01)(119,90)}
\SetColor{Magenta}
\Curve{(27.5,15.25)(45.4,14.89)(83.4,12.56)(119,11.07)}
\SetColor{BrickRed}
\SetColor{Black}
\SetScale{72}

\Text(-8,0)[]{\scriptsize $0$}
\Text(-8,18)[]{\scriptsize $2$}
\Text(-8,36)[]{\scriptsize $4$}
\Text(-8,54)[]{\scriptsize $6$}
\Text(-8,72)[]{\scriptsize $8$}

\Text(12,74)[l]{\scriptsize $\beta=0.375$}

\LinAxis(2.5,0)(2.5,1.25)(5,2,.040,0,.0100)

\SetOffset(-200,-260)
\SetWidth{.001}
\LinAxis(0,0)(0,1.25)(5,2,-.040,0,.0100)
\LogAxis(0,0)(2.5,0)(1.5,.040,5,.0100)
\SetScale{1}
\Vertex(45.4,31.95){1}
\Curve{(45.4,11.88)(45.4,52.02)}
\Vertex(83.4,32.04){1}
\Curve{(83.4,19.44)(83.4,44.64)}
\Vertex(119,55.98){1}
\Curve{(119,24.3)(119,87.66)}
\SetColor{Magenta}
\Curve{(27.5,1.800)(45.4,1.070)(83.4,.8500)(119,.7000)}
\SetColor{BrickRed}
\SetColor{Black}
\SetScale{72}
 
\Text(-8,0)[]{\scriptsize $0$}
\Text(-8,18)[]{\scriptsize $2$}
\Text(-8,36)[]{\scriptsize $4$}
\Text(-8,54)[]{\scriptsize $6$}
\Text(-8,72)[]{\scriptsize $8$}
 
\Text(12,74)[l]{\scriptsize $\beta=0.175$}
\LinAxis(2.5,0)(2.5,1.25)(5,2,.040,0,.0100)
 
\SetOffset(-200,-170)
\SetWidth{.001}
\LinAxis(0,0)(0,1.25)(5,2,-.040,0,.0100)
\LogAxis(0,0)(2.5,0)(1.5,.040,5,.0100)
\SetScale{1}
\Vertex(116,80){1}
\SetColor{Magenta}
\Curve{(112,72)(120,72)}
\SetColor{BrickRed}
\SetColor{Black}
\Vertex(27.5,34.92){1}
\Curve{(27.5,12.78)(27.5,57.06)}
\Vertex(45.4,33.21){1}
\Curve{(45.4,17.73)(45.4,48.69)}
\SetColor{Magenta}
\Curve{(27.5,.1300)(45.4,.1100)}
\SetColor{BrickRed}
\SetColor{Black}
\SetScale{72}
 
\Text(-8,0)[]{\scriptsize $0$}
\Text(-8,18)[]{\scriptsize $2$}
\Text(-8,36)[]{\scriptsize $4$}
\Text(-8,54)[]{\scriptsize $6$}
\Text(-8,72)[]{\scriptsize $8$}
\rText(-21,72)[][l]{\scriptsize $F_{2}^{\pom}$}
 
\Text(133,80)[l]{\scriptsize\em H1 Data}
\Text(133,72)[l]{\scriptsize\em Two gluon}
 
\Text(12,74)[l]{\scriptsize $\beta=0.065$}
 
\LinAxis(2.5,0)(2.5,1.25)(5,2,.040,0,.0100)
\LogAxis(0,1.25)(2.5,1.25)(1.5,-.040,5,.0100)
 
\SetOffset(26,-440)
\SetWidth{.001}
\LinAxis(0,0)(0,1.25)(5,2,-.040,0,.0100)
\LinAxis(0,0)(2.5,0)(5,2,.040,0,.0100)
\SetScale{1}
\Vertex(31.5,55.98){1}
\Curve{(31.5,24.3)(31.5,87.66)}
\Vertex(67.5,76.05){1}
\Curve{(67.5,44.01)(67.5,90)}
\Vertex(117,74.43){1}
\Curve{(117,38.88)(117,90)}
\SetColor{Magenta}
\Curve{(31.5,.7000)(67.5,11.07)(117,65.13)}
\SetColor{BrickRed}
\SetScale{72}
\SetColor{Black}
\Text(0,-7)[]{\scriptsize $0$}
\Text(36,-7)[]{\scriptsize $0.2$}
\Text(72,-7)[]{\scriptsize $0.4$}
\Text(108,-7)[]{\scriptsize $0.6$}
\Text(144,-7)[]{\scriptsize $0.8$}
\Text(180,-7)[]{\scriptsize $1$}
\Text(165,-17)[]{\scriptsize $\beta$}
\Text(90,-29)[]{\scriptsize $(b)$}
 
\Text(-8,0)[]{\scriptsize $0$}
\Text(-8,18)[]{\scriptsize $2$}
\Text(-8,36)[]{\scriptsize $4$}
\Text(-8,54)[]{\scriptsize $6$}
\Text(-8,72)[]{\scriptsize $8$}
 
\Text(12,74)[l]{\scriptsize $Q^2=50\,{\rm GeV}^2$}
\LinAxis(2.5,0)(2.5,1.25)(5,2,.040,0,.0100)

\SetOffset(26,-350)
\SetWidth{.001}
\LinAxis(0,0)(0,1.25)(5,2,-.040,0,.0100)
\LinAxis(0,0)(2.5,0)(5,2,.040,0,.0100)
\SetScale{1}
\Vertex(31.5,32.04){1}
\Curve{(31.5,19.44)(31.5,44.64)}
\Vertex(67.5,34.92){1}
\Curve{(67.5,23.85)(67.5,45.99)}
\Vertex(117,51.12){1}
\Curve{(117,31.41)(117,70.83)}
\SetColor{Magenta}
\Curve{(31.5,.8500)(67.5,12.56)(117,79.00)}
\SetColor{BrickRed}
\SetColor{Black}
\SetScale{72}
 
\Text(-8,0)[]{\scriptsize $0$}
\Text(-8,18)[]{\scriptsize $2$}
\Text(-8,36)[]{\scriptsize $4$}
\Text(-8,54)[]{\scriptsize $6$}
\Text(-8,72)[]{\scriptsize $8$}
 
\Text(12,74)[l]{\scriptsize $Q^2=25\,{\rm GeV}^2$}
\LinAxis(2.5,0)(2.5,1.25)(5,2,.040,0,.0100)

\SetOffset(26,-260)
\SetWidth{.001}
\LinAxis(0,0)(0,1.25)(5,2,-.040,0,.0100)
\LinAxis(0,0)(2.5,0)(5,2,.040,0,.0100)
\SetScale{1}
\Vertex(11.7,33.21){1}
\Curve{(11.7,17.73)(11.7,48.69)}
\Vertex(31.5,31.95){1}
\Curve{(31.5,11.88)(31.5,52.02)}
\Vertex(67.5,42.39){1}
\Curve{(67.5,29.7)(67.5,55.08)}
\Vertex(117,34.02){1}
\Curve{(117,14.67)(117,53.37)}
\SetColor{Magenta}
\Curve{(11.7,.1100)(31.5,1.070)(67.5,14.89)(108,81)}
\SetColor{BrickRed}
\SetColor{Black}
\SetScale{72}
 
\Text(-8,0)[]{\scriptsize $0$}
\Text(-8,18)[]{\scriptsize $2$}
\Text(-8,36)[]{\scriptsize $4$}
\Text(-8,54)[]{\scriptsize $6$}
\Text(-8,72)[]{\scriptsize $8$}
 
\Text(12,74)[l]{\scriptsize $Q^2=12\,{\rm GeV}^2$}
\LinAxis(2.5,0)(2.5,1.25)(5,2,.040,0,.0100)

\SetOffset(26,-170)
\SetWidth{.001}
\LinAxis(0,0)(0,1.25)(5,2,-.040,0,.0100)
\LinAxis(0,0)(2.5,0)(5,2,.040,0,.0100)
\SetScale{1}
 
\Vertex(116,80){1}
\SetColor{Magenta}
\Curve{(112,72)(120,72)}
\SetColor{BrickRed}
\SetColor{Black}
\Vertex(11.7,34.92){1}
\Curve{(11.7,12.78)(11.7,57.06)}
\SetColor{Magenta}
\Curve{(11.7,.1300)(31.5,1.800)(67.5,15.25)(108,90)}
\SetColor{BrickRed}
\SetColor{Black}
\SetScale{72}
 
\Text(-8,0)[]{\scriptsize $0$}
\Text(-8,18)[]{\scriptsize $2$}
\Text(-8,36)[]{\scriptsize $4$}
\Text(-8,54)[]{\scriptsize $6$}
\Text(-8,72)[]{\scriptsize $8$}
\rText(-21,72)[][l]{\scriptsize $F_{2}^{\pom}$}
 
\Text(133,80)[l]{\scriptsize\em H1 Data}
\Text(133,72)[l]{\scriptsize\em Two gluon}
 
\Text(12,74)[l]{\scriptsize $Q^2=8.5\,{\rm GeV}^2$}
\LinAxis(2.5,0)(2.5,1.25)(5,2,.040,0,.0100)
\LinAxis(0,1.25)(2.5,1.25)(5,2,-.040,0,.0100)

\end{picture}
\vspace*{6.3in}
\caption[Structure function $\ftwopom$ for the two gluon exchange
model.]  {{\it The two-gluon exchange model is fitted to virtuality
cut data~\cite{Phillips:1995jpp} for $\ftwopom$, with parameters as in
Fig.\,\ref{fig:erdl108multigraph}.}
\label{fig:2gluef2d2108}}
\end{center}
\end{figure}

We have also considered the diffractive scattering models for the case of
a larger pomeron intercept, $\intercept=1.2$, which corresponds to the
pomeron intercept favoured in recent H1 and ZEUS
analyses\,\cite{Adloff:1997sc,Breitweg:1998gc}. This choice of
intercept does not provide a significantly different fit, as can be
seen in Table\,\ref{table:summary}. This indicates that the present
data are not sufficiently precise to determine the intercept
accurately. As may be seen from Table\,\ref{table:summary}, the fits
are also reasonable for the choice $\intercept=1$. This corresponds to
the use of the $\xpom$ dependence as given by the graphs themselves, a
procedure adopted in\,\cite{Nikolaev:1992et,Hautmann:1998xn}.

\begin{table}[htp]
\centering
\begin{tabular}{|l|ccc|}
\hline
&  &  &\\ [-6pt]
& \multicolumn{3}{c|}{$\chi^2/{\rm dof}$} \\ 
&  &  &\\ [-6pt]\cline{2-4}
&  &  &\\ [-6pt]
& ~~$\alpha_{\pom}(0)=1.08$~~ & ~~$\alpha_{\pom}(0)=1.2$~~& ~~$\alpha_{\pom}(0)=1$~~ \\
&  &  &\\[-6pt] \hline
&  &  &\\[-6pt]
E-R: $f(k^2)=\mbox{\Large $\sqrt{\frac{\Lambda^2}{\Lambda^2-k^2}}$}$ &
37/41 & 54/41 & 30/41\\ 
&  &  &\\ [-6pt]
&  &  &\\ [-6pt]
D-L: $f(k^2)=\mbox{\Large $\frac{\Lambda^2}{\Lambda^2-k^2}$}$ & 102/41 &  97/41 &179/41\\ 
&  &  &\\ [-6pt]
&  &  &\\ [-6pt]
Single gluon:    &  64/41  &  57/41 &69/41\\ 
&  &  &\\ [-6pt]
&  &  &\\ [-6pt]
Two gluon (Diehl):  &  137/41   & 136/41 &137/41\\ 
&  &  &\\ [-6pt]
&  &  &\\ [-6pt]
Scalar (VBLY):  &   48/41   &  61/41 &40/41\\ 
&  &  &\\ \hline
\end{tabular}
\caption[Result of the fit of the various colour singlet exchange
models to~$F_{2}^{D(3)}$ structure function data.]{{\it Results of the
fits of
various colour-singlet exchange models to the~$F_{2}^{D(3)}$ structure
function data from~\cite{Phillips:1995jpp}.}}
\label{table:summary}
\end{table}

\section{Future Experiments}

As has been discussed previously\,\cite{Ellis:1996cg} and was also
discussed above, H1 and ZEUS analyses in which only events with very
large pseudo-rapidity gaps were selected are not described by the
resolved-coupling picture as previously formulated~\cite{Ingelman:1985ns},
as the cuts force
the partons coupling to the pomeron to have large virtualities. Thus
these experiments provide a very strong test of models of the
colourless component of the pomeron coupling to the virtual quark. 
Here we wish to stress that these tests can be made more
stringent by relatively straightforward measurements.

For a given pseudo-rapidity cut, one can calculate the region of
parameter space in $\qsq$, $\beta$ and $\xpom$ in which one would
expect to see strong virtuality constraints due to the cut, using the
results presented in Appendices\,A.1 and\,A.2. There are a number of
ways one can use this information to test these ideas. The dependence
of the exchanged parton virtuality on pseudo-rapidity cuts and
kinematic parameters could best be studied by examining two samples
of data. The first should be chosen with a relatively strong
pseudo-rapidity cut, in a region where the cuts force $|k^2|\gtrsim$ a
few GeV$^2$. The second set should be chosen to differ from the first
only in the strength of the pseudo-rapidity cut. For a stronger cut,
one would expect to see a relative reduction in the extracted
diffractive structure function due to there being less phase space
available for the scattering interaction. This is a model-independent
effect, and does not require knowledge of the overall normalization.

The size of the reduction is a direct indication of the magnitude of
the colourless component of the pomeron coupling to the virtual quark
system. The magnitude of this reduction is predicted 
differently by the various
models of diffractive scattering, and hence offers a sensitive way to
discriminate between them. At present, we know of only one such study
with data chosen with two different pseudo-rapidity
cuts\,\cite{Adloff:1997sc}, but the virtuality constraints used are
insufficient to provide a significant reduction in phase space.

Finally, as noted above, the $\qsq$, $\beta$ and $\xpom$ dependence of
the pseudo-rapidity cut provides a further sensitive test of the
models. At present, the data available for such a
study\,\cite{Phillips:1995jpp} have rather poor statistics, but can
already discriminate between several of the models presented
here. More accurate data would be extremely useful in refining the
selection between models.

\section{Summary and Conclusions}

\label{sect:summary}

In this paper we have re-examined the information that can be obtained
from the pseudo-rapidity gap dependence of diffractive deep-inelastic
scattering. The main conclusion, up to graphs involving soft parton
emission, is that a strong pseudo-rapidity cut requires that the
resolved partonic component of the Pomeron must be colourless, i.e.,
the coupling of the Pomeron for partonic states of high virtuality or
the coupling is via colour-singlet states. This may be achieved either
by postulating a direct coupling with an associated form factor or by
modelling it via multi-gluon and multi-quark states. Imposing the
pseudo-rapidity cut provides a way to measure this component in the
coupling of the Pomeron.

Analysis of the available data with strong pseudo-rapidity cuts shows
that the cross section is relatively insensitive to the cut, implying
that the direct or multi-parton component of the pomeron coupling makes a
significant contribution. The fit assuming a direct $C=+1$ vector-like
coupling to quarks is sensitive to the form factor describing the
dependence on the virtuality of the quark. An excellent fit is
obtained using the hard E-R form factor, whereas the softer D-L form is
not consistent with the data. A good fit is also obtained for the scalar
pomeron case. A marginally acceptable fit is also obtained for the
case that diffractive scattering is described by the leading-twist
single-gluon component with colour dressing due to soft gluons. This
contribution necessarily has the pomeron momentum carried by the
single gluon. If this is indeed the source of the large
pseudo-rapidity gap events, the fact that these events are a sizeable
part of the full diffractive deep-inelastic events implies that a
significant partonic component of the Pomeron has all of its momentum
carried by a single gluon.

This is very interesting in terms of the recent H1 fit of
diffractive structure function data\,\cite{Adloff:1997sc} using the
diagrams of Figs.\,\ref{fig:dijet} and\,\ref{fig:3jets}, which
favoured a strongly peaked gluon component with one very hard
gluon. However, the H1 calculation considered only the resolved-coupling
component of the pomeron, which we argue here is unable to
describe data for which the pseudo-rapidity cut imposes strong
virtuality constraints. Further, it was suggested in~\cite{Ellis:1996cg}
that the contribution from Fig.\,\ref{fig:2glue} might have
significant scaling violations which, if included in the H1 fit, might
enable a less extreme pomeron structure to be fitted to the observed
diffractive structure function scaling violations. We also note that
there are alternative fits to the H1 data in a two-gluon exchange
model~\cite{Bartels:1998ea}.

Perhaps the most significant feature of the comparison of the models
with data is the failure of the two-gluon model of Diehl to fit the
observed $\beta$ dependence. The reason for this is that, with a large
pseudo-rapidity gap cut, there is a strong bound on the lowest $\ptsq$
available. Imposing this, the model predicts a $\beta^3$ dependence for
the structure function, which is in strong disagreement with experiment.
One may
ask whether this feature persists in all two-gluon exchange models, as
Diehl assumes a particular form for the ``non-perturbative'' gluon
propagators. In the case of the more phenomenological model
of\,\cite{Bartels:1998ea} it is easy to check that the $\beta^3$
behaviour should also apply to the $q\bar{q}$ longitudinal and
transverse components. The only remaining term in the model which
might give a better $\beta$ dependence for the case of a strong
$\ptsq$ lower bound is the $q\bar{q}g$ component. Unfortunately we do
not know this dependence at present.

Given the sensitivity of various models of diffractive processes in
deep-inelastic scattering to the pseudo-rapidity cut, it would be of
great interest to obtain improved statistics for such processes. This
may allow us to distinguish between the direct coupling and the
leading-twist models.

\section*{Acknowledgments}

J.W. wishes to thank Julian Phillips for helpful discussions and to
gratefully acknowledge funding by a Commonwealth Scholarship.

\renewcommand{\theequation}{A.\arabic{equation}} \setcounter{equation}{0} %
\renewcommand{\thesection}{A.\arabic{section}} \setcounter{section}{0}
\appendix

\section*{Appendix\,A}
\label{app:a}

\subsection*{A.1~~Constraints on Parton Virtualities in Dijet Production}

Here we consider the effect of pseudo-rapidity cuts on the virtuality
of the exchanged quark in~Fig.\,\ref{fig:dijet}. This calculation was
originally reported in~\cite{Ellis:1996cg}, which also includes an
analysis of the data shown in~Table\,\ref{table:ervirtuality}. In this
Appendix we present the derivation of this result. The outline is as
follows: we start with pseudo-rapidity defined in terms of the polar
angle in the laboratory (LAB) frame, and then boost to the
photon-pomeron centre-of-mass (CMS) frame, which is also the CMS frame
of the hadronic jets in the diffractively-produced system~$X$. From
this one can relate pseudo-rapidity to struck quark virtuality in terms of
the kinematic invariants, and therefore determine the minimum
virtualities,~$k^2_{\rm min}$, implied by the pseudo-rapidity
cuts,~$\eta_{\rm max}$, for each set of parameters,~$\beta$,~$Q^2$,
and~$x_{I\!\!P}$.

\begin{figure}[htp]
\begin{center} \begin{picture}(180,130)(-20,0)
\SetScale{0.8}
\Text(9,110)[]{\scriptsize $p_{e}$}
\Text(85,117)[]{\scriptsize $p_{e}^{\prime}$}
\Text(113,83)[]{\scriptsize $\pa$}
\Text(154,64)[]{\Huge \}}
\Text(164,64)[]{$X$}
\Text(113,45)[]{\scriptsize $\pb$}
\Text(60,67)[]{\scriptsize $k$}
\Text(20,6)[]{\scriptsize $P$}
\Text(112,6)[]{\scriptsize $P^{\prime}$}
%
\Line(5,128)(60,128)
\Line(60,128)(105,155)
\Line(84,103)(130,103)
\Line(84,58)(130,58)
\ArrowLine(84,103)(84,58)
\Line(30,13)(130,13)
\Photon(60,128)(84,103){4}{3.5}
\ZigZag(84,58)(84,13){-4}{4.5}
\Vertex(60,128){1}
\Vertex(84,103){1}
\Vertex(84,58){1}
\Vertex(84,13){1}
\end{picture}
\caption[Momentum assignments for dijet production.]{{\it Momentum
assignments for dijet production.}\label{fig:dijetmomenta}}
\end{center}
\end{figure}
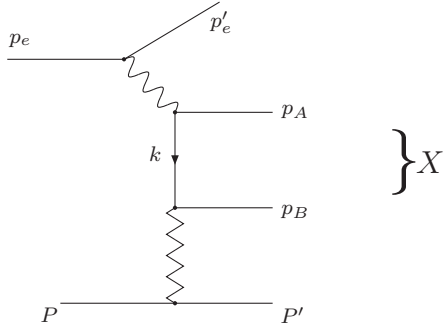

We consider diffractive~$e-P$ deep-inelastic scattering via dijet
production in the HERA LAB frame:

\begin{equation}
e(p_{e}) + P(P) \rightarrow e(p_{e}^{\prime}) + X(X_{\rm lab}) + P(P^{\prime}),
\end{equation}

\noindent where the momenta of the particles are shown in brackets,
and~$\mxsq$ is the invariant mass squared of the diffractive system
$X$ composed of the two outgoing jets, as shown in
Fig.\,\ref{fig:dijetmomenta}. In the LAB frame,

\begin{eqnarray}
p_{e} & = & (E_{e},\,0.\,0,\,-E_{e}) ~~~~~~~~~~~~~~~~~~~
E_{e}=27.5\,{\rm GeV}\hspace*{1.5cm} 
\nonumber \\
P & = & (E_{P},\,0.\,0,\,E_{P}) ~~~~~~~~~~~~~~~~~~~~ E_{P}=820\,{\rm GeV} 
\nonumber \\
p_{e}^{\prime} & = &
(E_{e}^{\prime},\,E_{e}^{\prime}\sin\theta_{\rm lab},\,0,\,E_{e}^{\prime}\cos
\theta_{\rm lab}) \nonumber \\
q &=& (E_{e}-E_{e}^{\prime},\,-E_{e}^{\prime}\sin\theta_{\rm lab},\,0,\,
-E_{e}-E_{e}^{\prime}\cos\theta_{\rm lab}),
\end{eqnarray}

\noindent
and we parameterize the lower quark momentum generally by

\be
\pb=(l,\,l\sin\theta^{\prime}_{\rm lab}\cos\phi_{\rm lab},\,
l\sin\theta^{\prime}_{\rm lab}\sin\phi_{\rm lab},\,l\cos\theta^{\prime}_{\rm lab}).
\ee

\noindent
Here we have assumed that the pomeron is emitted in the proton
direction and carries a fraction~$\xpom$ of the proton initial momentum.

The following relations are useful:

\begin{equation}
\qsq=-q^2=-(\pe-\pep)^2=2\pe\cdot\pep=2E_{e}E_{e}^{\prime}(1+\cos\theta_{\rm lab})
\end{equation}

\noindent
and

\bea
W^2=(P+q)^2\Rightarrow W^2+\qsq =2P\cdot q &=&
2E_{P}(E_{e}-E_{e}^{\prime}+E_{e}+E_{e}^{\prime}\cos\theta_{\rm lab})\nonumber \\
&=& 2E_{P}[2E_{e}-E_{e}^{\prime}(1-\cos\theta_{\rm lab})].
\eea

\noindent
Hence

\bea
E_{e}^{\prime}(1+\cos\theta_{\rm lab})&=&\frac{\qsq}{2E_{e}} \label{eq:A}
\\
E_{e}^{\prime}(1-\cos\theta_{\rm lab})&=&2E_{e}-
\frac{\qsq+W^2}{2E_{P}}. \label{eq:B}
\eea

\noindent
Therefore, adding (\ref{eq:A}) and (\ref{eq:B}), we get

\begin{equation}
2E_{e}^{\prime}=2E_{e}+ \frac{\qsq}{2E_{e}} - \frac{\qsq}{2E_{P}\beta \xpom
},
\end{equation}

\noindent
and, subtracting,

\begin{equation}
2E_{e}^{\prime}\cos\theta_{\rm lab}=-2E_{e}+ \frac{\qsq}{2E_{e}} + \frac{\qsq}{
2E_{P}\beta \xpom}.
\end{equation}

\noindent
We can therefore write the momentum of the diffractive system as

\bea
X_{\rm lab}
& =&(\xpom E_{P}+E_{e}-E_{e}^{\prime},\,
-E_{e}^{\prime}\sin\theta_{\rm lab},\,0,\,
\xpom E_{P}-E_{e}-E_{e}^{\prime}\cos\theta_{\rm lab}),\nonumber \\
&=&\left(\xpom E_{P}+\frac{\qsq}{2}\left[\frac{1}{2 E_{P}\beta x
_{\pom}}-\frac{1}{2E_{e}}\right],\, -\left[\qsq\left
(1-\frac{\qsq}{2E_{P} E_{e}\beta x
_{\pom}}\right)\right]^{\frac{1}{2}}, \right. \nonumber \\
&& \hspace*{4cm} \left. 0,\xpom E_{P}-\frac{\qsq}{2}\left[\frac{1}{2
E_{P}\beta \xpom}+\frac{1}{2E_{e}}\right]\right).
\eea

\noindent
In the CMS frame, the momentum of the diffractive system is

\begin{equation}
X_{\rm cms}=(M_{X},\,0,\,0,\,0).
\end{equation}

\vspace*{0.3cm}

We now carry out two Lorentz boosts to get from the LAB frame to the jet-jet
CMS frame, 
%
%
seeking to remove the spatial components of $X_{\rm lab}$.
The first boost we take along the~$z$~axis, with boost
parameter~$\beta_{1}$. The required boost satisfies
$X^{z}_{\rm lab}-\beta_{1}X^{0}_{\rm lab}=0$, giving

\be
\beta_{1}
= \frac{ -\left[ \frac{\qsq}{2}\left(\frac{1}{2E_{P}\beta \xpom}
+\frac{1}{2E_{e}}\right)-\xpom E_{P} \right]}
{\frac{\qsq}{2}\left(\frac{1}{2E_{P}\beta \xpom}
-\frac{1}{2E_{e}}\right)+\xpom E_{P}}.
\label{eq:beta1}
\ee

\noindent
After this first boost, we have

\bea
X^{0}_{\rm lab}\rightarrow X^{\prime\,0}
&=&\frac{1}{\sqrt{1-\beta_{1}^2}}\left[\xpom E_{P}+
\frac{\qsq}{2}\left(\frac{1}{2E_{P}\beta\xpom}-\frac{1}{2E_{e}}
\right)-\right. \nonumber\\
&&\beta_{1}\left\{\xpom E_{P}-\frac{\qsq}{2}
\left(\frac{1}{2E_{P}\beta\xpom}+\frac{1}{2E_{e}}\right)
\right\}\left.\right] \nonumber \\
&=& \frac{\sqrt{1-\beta_{1}^2}}{\beta_{1}}
\left[\xpom E_{P}-\frac{\qsq}{2}\left(
\frac{1}{2E_{P}\beta\xpom}+\frac{1}{2E_{e}}\right)\right], \nonumber \\
&&\nonumber \\
&X^{\prime\,x}=X^{X}_{\rm lab}.
\eea

\noindent
We now boost along the~$x$~direction, which requires a boost
parameter~$\beta_{2}$ satisfying

\be
X^{\prime\,x}-\beta_{2}X^{\prime\,0}=0,
\ee

\noindent
giving

\be
\beta_{2}
= \frac{\beta_{1}}{\sqrt{1-\beta_{1}^{2}}}\,\frac{\left[\qsq\left(1-
\frac{\qsq}{4E_{e}E_{P}\beta \xpom}\right)\right]^{\frac{1}{2}}} {\frac{
\qsq}{2}\left(\frac{1}{2E_{P}\beta \xpom}+\frac{1}{2E_{e}}
\right)-\xpom E_{P}}.  \label{eq:beta2}
\ee

\noindent 
These two boosts take us to the jet-jet CMS frame.

Now we consider the jet produced by the lower parton line in the LAB frame.
Consider the effect of the previous two boosts on the~$0^{{\rm th}}$
component of~$\pb$:

\begin{eqnarray}
\beta_{1}:~~\pb^{0} &\!\!\!\!\!\!
\rightarrow ~p^{\prime\,0}_{B}&\!\!= 
\frac{l}{\sqrt{1-\beta_{1}^{2}}}(1-\beta_{1}\cos\theta_{\rm lab}),\hspace*{0.5cm}
\nonumber \\
\beta_{2}:~~p^{\prime\,0}_{B} &\!\!\!\!\!\!
\rightarrow ~p_{B\,{\rm cms}}^{0}&\!\!= 
\frac{l}{\sqrt{1-\beta_{2}^{2}}}\left[\frac{(1-\beta_{1}\cos\theta_{\rm lab})}
{\sqrt{1-\beta_{1}^{2}}}-
\beta_{2}\sin\theta_{\rm lab}\cos\phi_{\rm lab}\right].\nonumber \\
&&
\end{eqnarray}

\noindent
Remembering that

\be
P_{\rm lab}=(E_{P},\,0,\,0,\,E_{P}),
\ee

\noindent
we see that the effect of the boosts on the~$0^{{\rm th}}$ component
of the proton initial momentum is

\begin{eqnarray}
\beta_{1}:~~~P^{0} &\rightarrow P^{\prime\,0}&=\frac{(1-\beta_{1})}
{\sqrt{1-\beta_{1}^{2}}}E_{P} =\sqrt{\frac{1-\beta_{1}}{1+\beta_{1}}}E_{P},
\nonumber \\
\beta_{2}:~~~P^{\prime\,0} &\rightarrow P^{0}_{\rm cms}& =\sqrt{
\frac{1-\beta_{1}}{1+\beta_{1}}}\frac{E_{P}}{\sqrt{1-\beta_{2}^{2}}}.
\end{eqnarray}

\noindent
For general on~shell 4-vectors~$p$ and~$q$, we have

\begin{equation}
p\cdot q = p^{0}\,q^{0}\,(1-\cos\theta),
\end{equation}

\noindent
where~$\theta$ is the angle between~$p$ and~$q$. So we have

\begin{eqnarray}
P\cdot \pb &= &P^{0}\,\pb^{0}\,(1-\cos\theta_{\rm lab})  \nonumber \\
&=& P_{\rm cms}^{0}\,p_{B\,{\rm cms}}^{0}\,(1-\cos\theta_{\rm cms}),
\end{eqnarray}

\noindent
where $\theta_{\rm cms}$ is the angle of the quark relative to the forward
proton direction in the CMS frame, and $\theta_{\rm lab}$ is the angle of
the quark relative to the forward proton direction in the LAB frame. Hence

\be
(1-\cos\theta_{\rm cms})= \frac{ (1+\beta_{1})(1-\beta_{2}^{2})\,
(1-\cos\theta_{\rm lab})}{1-\beta_{1}\cos\theta_{\rm lab}-
\sqrt{1-\beta_{1}^2}\,\beta_{2}
\sin\theta_{\rm lab}\cos\phi_{\rm lab}}.  \label{eq:angles}
\ee

\noindent
Thence the constraint follows: a cut on
pseudo-rapidity,~$\eta_{\rm max}$, is
equivalent to a cut on the lab angle,~$\theta_{\rm lab}^{\rm min}$, where

\begin{equation}
\eta=-\ln\tan\frac{\theta_{\rm lab}}{2}~~\Rightarrow~~~~~
\cos\theta_{\rm lab}^{\rm min} = \frac{1-e^{-2\eta_{\rm max}}}{1+e^{-2\eta_{\rm max}}}.
\end{equation}

\noindent
Thus

\begin{equation}
(1-\cos\theta_{\rm cms}^{\rm min}) = \frac{(1+\beta_{1})(1-\beta_{2}^{2})\frac{
2e^{-2\eta_{\rm max}}}{1+e^{-2\eta_{\rm max}}}} {1-\beta_{1}\frac{1-e^{-2\eta_{\rm max}}
}{1+e^{-2\eta_{\rm max}}}- \sqrt{1-\beta_{1}^2}
\beta_{2}\frac{2e^{-\eta_{\rm max}}}{1+e^{-2\eta_{\rm max}}}\cos\phi_{\rm lab}},
\end{equation}

\noindent from which it follows from~(\ref{eq:ksqdefn})
that a cut in pseudo-rapidity,~$\eta_{\rm max}$, corresponds to a minimum
virtuality,~$k^{2}_{\rm min}$, of the exchanged quark:

\be
-k^2_{\rm min}=\frac{\qsq}{2\beta}\,\frac{(1+\beta_{1})(1-\beta_{2}^{2})\frac{
2e^{-2\eta_{\rm max}}}{1+e^{-2\eta_{\rm max}}}} {1-\beta_{1}\frac{1-e^{-2\eta_{\rm max}}
}{1+e^{-2\eta_{\rm max}}}- \sqrt{1-\beta_{1}^2}
\beta_{2}\frac{2e^{-\eta_{\rm max}}}{1+e^{-2\eta_{\rm max}}}\cos\phi_{\rm lab}}.
\ee

\noindent
Since we do not know the azimuthal angle~$\phi_{\rm lab}$, we
choose~$\phi_{\rm lab}$ to minimize~$|k^{2}_{\rm min}|$, and still find
that for a large range of parameter space~$k^{2}_{\rm min}$ is
constrained to be much larger than~$1\,{\rm GeV}^{2}$, as seen in
Tables\,\ref{table:ervirtuality} and~\ref{table:phillipsvirtuality}.

This calculation demonstrates clearly that dijet production cannot
contribute to large pseudo-rapidity gap diffractive DIS for a wide range of
the parameters~$\beta$,~$\qsq$ and~$\xpom$.

\subsection*{A.2~~Constraints on Parton Virtualities for Production
of Three or More Jets}

For production of three or more jets, e.g., the processes
of~Fig.\,\ref{fig:3jets}, the situation is slightly more
complicated. However, here one also finds strong virtuality
constraints. It is easiest to consider multi-jet production
diffractive processes in the form

\begin{equation}
e(p_{e}) + P(P) \rightarrow e(p_{e}^{\prime}) + X_{1}(p_{A}) + X_{2}(p_{B})
+ P(P^{\prime}),
\end{equation}

\noindent
where~$X=X_{1}+X_{2}$ is the diffractive system. The system
$X_{2}$ is the hadronic jet produced by the final-state parton
coupled to the pomeron in a diagram such as~Fig.\,\ref{fig:3jets}, and
has squared mass~$M_{X_{2}}^2\lesssim 1\,{\rm GeV}^2$. The sum over
the jets formed from the other final-state partons in the process
is~$X_{1}$, and has squared mass~$M^2<M_{X}^2$. In the example of
three-jet production via boson-gluon fusion shown
in~Fig.\,\ref{fig:3jets}b, $X_1$~corresponds to the two upper 
final-state quark jets, whilst~$X_2$ represents the on-shell emitted
gluon. In this diagram, we are looking at constraints on the squared
four-momentum of the exchanged gluon.

\begin{figure}[htp]
\begin{center} \begin{picture}(180,130)(-20,0)
\SetScale{0.8}
\Text(9,110)[]{\scriptsize $p_{e}$}
\Text(85,117)[]{\scriptsize $p_{e}^{\prime}$}
\Text(113,83)[]{\scriptsize $p_A$}
\Text(139,83)[]{\large\}}
\Text(148,83)[]{$X_{1}$}
\Text(139,45)[]{\large\}}
\Text(148,45)[]{$X_{2}$}
\Text(113,45)[]{\scriptsize $p_B$}
\Text(60,67)[]{\scriptsize $k$}
\Text(20,6)[]{\scriptsize $P$}
\Text(112,6)[]{\scriptsize $P^{\prime}$}
\Line(84,84)(81,81)
\Line(84,84)(87,81)
\Line(5,128)(60,128)
\Line(60,128)(105,155)
\Line(84,103)(130,103) %
\Line(84,103)(130,100) %
\Line(84,103)(130,106) %
\Line(84,58)(130,58)
\Line(84,103)(84,58)
\Line(30,13)(130,13)
\Photon(60,128)(84,103){4}{3.5}
\ZigZag(84,58)(84,13){-4}{4.5}
\Vertex(60,128){1}
\Vertex(84,103){1}
\Vertex(84,58){1}
\Vertex(84,13){1}
\end{picture}
\caption[Momentum assignments for multi-jet production.]{{\it Momentum
assignments for multi-jet production.}\label{fig:multijetmomenta}}
\end{center}
\end{figure}
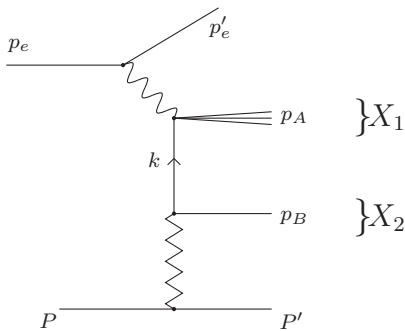

In the CMS frame, we parameterize the jet momenta by

\begin{eqnarray}
p_{A}& =&\gamma\frac{\mxsq}{2}\,
(\frac{2-\gamma}{\gamma},\,-\sin\theta_{\rm
cms},\,0,\,-\cos\theta_{\rm cms})  \nonumber \\
p_{B}& =&\gamma\frac{\mxsq}{2}\,
(1,\,\sin\theta_{\rm cms},\,0,\,\cos\theta_{\rm cms}),
\end{eqnarray}

\noindent
where the mass squared of system~$X_{1}$ is $(1-\gamma)\mxsq$, and
$0<\gamma\leq 1$. The case~$\gamma=1$ corresponds to dijet
production. The pomeron momentum in this frame is

\begin{equation}
P_{I\!\!P}=(E,\,0,\,0,\,E),
\end{equation}

\noindent
where

\begin{equation}
E=\frac{Q^2+M_{X}^2}{2M_{X}}.
\end{equation}

\noindent
The momentum of the exchanged parton is given by

\begin{equation}
k=p_{A}-q=P_{I\!\!P}-p_{B}
\end{equation}

\noindent
and therefore the virtuality of the exchanged parton is

\begin{equation}
k^2= -\gamma\frac{Q^2+M_{X}^2}{2}(1-\cos\theta_{\rm cms}).  \label{eq:gammakqsqdefn}
\end{equation}

\noindent
We see that the constraint from the~$\eta_{\rm max}$ cuts, derived in
Appendix\,A.1, can in theory be evaded by having~$\gamma$ small.

Let us first consider the case $\theta_{\rm cms}>90\degg$, in which case 
the condition that the virtuality of the parton be low (say $\leq
1\,{\rm GeV}^2$) requires

\be
\gamma\leq1\,{\rm GeV}^2\cdot\frac{2}{\qsq+\mxsq}=\frac{2}{k^2_{\rm
max}\,({\rm GeV}^2)}.
\ee

\noindent
From Table\,\ref{table:phillipsvirtuality}, we see that for the case
that the pseudo-rapidity cut imposes a significant constraint on
$\kmin$, $\kmin>6\,{\rm GeV}^2$, say, $\gamma\leq1/33$. This
corresponds effectively to a two-jet configuration, because the soft
parton
carrying less than $1/33$ of the Pomeron momentum will not be observed
as a jet. The other parton is constrained by the pseudo-rapidity cut
to carry essentially all of the Pomeron momentum. As discussed in the
text, this contribution is included in the model corresponding to
leading twist, but not in the models of direct coupling. The non-soft
contributions which are genuinely multi-jet processes may be shown to
require high virtuality similar to the two-jet bound. To see this, we note
that, from~(\ref{eq:gammakqsqdefn}), the pseudo-rapidity cuts give
us

\begin{eqnarray}
-k^2_{\gamma<1}\geq-k^2_{\rm min}&=&\gamma\frac{Q^2+M_{X}^2}{2}
(1-\cos\theta_{\rm cms}^{\rm min}) \nonumber \\
&=& -\gamma\, k^2_{\gamma=1},  \label{eq:gammakqsqlimit}
\end{eqnarray}

\noindent
where~$k^2_{\gamma=1}$ is the minimum virtuality of the exchanged
quark in the dijet production diagram of Fig.\,\ref{fig:dijet} allowed
by the experimental cuts.

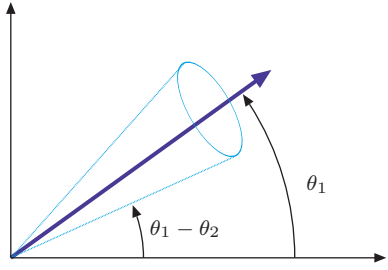
\begin{figure}[htp]
\centering
\begin{picture}(180,130)(0,0)
\SetScale{1}
\SetColor{BlueViolet}
\SetWidth{1.5}
\LongArrow(0,0)(97,70)
\SetWidth{0.5}
\SetColor{Cerulean}
\DashLine(0,0)(65,73){0.3}
\DashLine(0,0)(84,38){0.3}
\Oval(75,56)(20,8)(30)
\SetColor{Black}
\LongArrow(0,0)(140,0)
\LongArrow(0,0)(0,95)
\LongArrowArc(0,0)(107,0,35.2)
\LongArrowArc(0,0)(50,0,23.4)
\Text(116,28)[]{\scriptsize $\thetaone$}
\Text(66,11)[]{\scriptsize $\thetaone-\thetatwo$}
\end{picture}
\caption[Pseudo-rapidity definition at parton and at hadron level.]
{{\it Pseudo-rapidity definition at the parton and hadron
levels.}}\label{fig:angles}
\end{figure}

The opening angle of a soft jet will clearly be much greater than for
a harder jet. We quantify this statement below. The angles discussed
here are shown in Fig.\,\ref{fig:angles}. In the HERA LAB frame,
$\thetaone$ is the angle between the final-state parton which
hadronizes to produce the observed jet, and $\thetatwo$ is the opening
angle of the hadronic jet. Therefore $\thetaone-\thetatwo$ is the
angle between the edge of the hadronic jet and the proton direction,
in terms of which the pseudo-rapidity gap is defined.

\noindent
To estimate the opening angle of a soft jet, we assume that a jet
produced by a parton of energy $E_j$ has a cone radius of 0.5 to 1 unit
of pseudo-rapidity, and an opening angle of $\thetatwo$. Boosting to a frame
in which the parton has energy $\gamma\,E_j$, we can find the opening
angle, $\thetatwop$, of the corresponding jet in terms of
$\thetatwo$. This is the relation between angles that we shall assume
for a soft jet in the HERA LAB frame. Starting with a massless
parton with energy $E_j$, we then boost to a frame in which it has
energy $\gamma\,E_j$:

\be
(E_j,\,0,\,0,\,E_j) ~~\to~~
\gamma\,(E_j,\,0,\,0,\,E_j)
\ee

\noindent
If we denote the boost parameters by $\gammast$ and $\betast$, where
$\gammast=1/\sqrt{1-\betast^2}$, then we have

\be
\gamma\,E_j \rightarrow
\gammast(E_j-\betast\,E_j) ~~\Rightarrow~~
\gamma=\gammast(1-\betast).
\ee

\noindent
The corresponding transformation of angles is

\bea
\tan\thetatwop &=& \frac{1}{\gammast}\,
\frac{\sin\thetatwo}{\cos\thetatwo-\betast} \nonumber \\
 &=& \frac{1}{\gammast}\,\frac{\sin\thetatwo/\cos\thetatwo}
{1-\betast/\cos\thetatwo} \nonumber \\
 &\geq& \frac{1}{\gammast}\,\frac{\tan\thetatwo}{1-\betast}
\nonumber \\
 &=& \frac{1}{\gamma}\,\tan\thetatwo \label{eq:tantheta}
\eea

\noindent
Further, from (\ref{eq:tantheta}), we have

\be
\tan\thetatwop\geq\frac{1}{\gamma}\tan\thetatwo\geq\frac{1}{\gamma}
2\tan\frac{\thetatwo}{2}
\ee

\noindent
i.e.,

\be
\tan\frac{\thetatwop}{2}\geq\frac{1}{\gamma}\tan\frac{\thetatwo}{2}\,
(1-\tan^2\frac{\thetatwop}{2})
\ee

\vspace*{0.3cm}

\noindent
To see what sort of bound we can put on
$(1-\tan^2\frac{\thetatwop}{2})$, we assume that $\thetaone\leq
90^{\circ}$, i.e., the jet is produced in the forward hemisphere.

The large pseudo-rapidity cut data from~\cite{Phillips:1995jpp}
was taken with an experimental cut on pseudo-rapidity of
$\etamax=1.8$, which corresponds to a minimum angle in the HERA LAB
frame with no hadronic activity of \mbox{$\theta_{\rm min}^{\rm lab} \approx
18^{\circ}$}. Therefore

\be
\thetatwop\leq\thetaone-18\degg~~\Rightarrow~~
1-\tan^2\frac{\thetatwop}{2}\geq 1-\tan^2 36\degg \approx 0.47.
\ee

\noindent
Thus

\be
\tan\frac{\thetatwop}{2}\geq\frac{0.47}{\gamma}\tan\frac{\thetatwo}{2}.
\label{eq:tangents}
\ee






\noindent
Finally, 
%
%
pseudo-rapidity cuts are defined at the experimental level by seeking
events with an angle $\thetaone-\thetatwo$ which satisfies the
pseudo-rapidity cut $\eta_{\rm max}$:

\bea
\etamaxexp & \geq & -\ln\tan\left(\frac{\thetaone-\thetatwo}{2}\right)
\nonumber \\
&\geq&-\ln(\tan\frac{\thetaone}{2}-\tan\frac{\thetatwo}{2}) \nonumber \\
&=&-\ln\tan\frac{\thetaone}{2}-
\ln\left(1-\frac{\tan\frac{\thetatwo}{2}}{\tan\frac{\thetaone}{2}}\right)
\nonumber \\
~ \nonumber \\
&\Rightarrow&-\ln\tan\frac{\thetaone}{2} \leq \etamaxexp
+\ln\left(1-\frac{\tan\frac{\thetatwo}{2}}{\tan\frac{\thetaone}{2}}\right).
\eea

\noindent
At the theoretical quark-parton level, we deal with the angle $\thetaone$,
and assume a cone
radius of 0.5 to 1 unit of pseudo-rapidity. Therefore

\be
-\ln\tan\frac{\thetaone}{2}\leq\etamaxth \approx \etamaxexp-(\ha\to1),
\ee

\noindent
giving

\bea
\tan\frac{\thetatwo}{2}= (^{0.39}_{0.63})\tan\frac{\thetaone}{2}, \label{eq:moreangles}
\eea

\noindent
where the factor 0.39 is for the case where one assumes a cone radius
of $\ha$, and the lower number is for unit radius.

\vspace*{0.3cm}

Putting this all together, we can determine the bound on multi-jet
production in terms of the dijet limit calculated in Appendix\,A.1.
Using (\ref{eq:angles}), we can write

\be
(1-\cos\theta\cms)=f\,(1-\cos\theta\lab), \label{eq:angles2}
\ee

\noindent
where $f$ is a complicated function of the LAB angles and boost
parameters. We therefore have:

\bea
k^2_{\gamma<1}&=& \gamma  \frac{\qsq+\mxsq}{2}\,(1-\cos\thetaone\cms) \nonumber \\
&=&\gamma \frac{\qsq+\mxsq}{2}\,f\, (1-\cos\thetaone\lab) \nonumber \\
&=& \gamma \frac{\qsq+\mxsq}{2}\,f\, \tan\frac{\thetaone\lab}{2}
{\sin\thetaone\lab} \nonumber \\
&\geq& \gamma \frac{\qsq+\mxsq}{2}\,f\,
\tan\frac{\thetatwop\lab}{2}{\sin\thetaone\lab} \nonumber \\
&\geq& 0.47 \frac{\qsq+\mxsq}{2}\,f\,\tan\frac{\thetatwo\lab}{2}
{\sin\thetaone\lab} \nonumber \\
&=& 0.47 \frac{\qsq+\mxsq}{2}\,f\, (^{0.39}_{0.63})\,\tan\frac{\thetaone\lab}{2}
{\sin\thetaone\lab} \nonumber \\
&\approx&  (^{0.2}_{0.3})\,\frac{\qsq+\mxsq}{2}f\,\tan\frac{\thetaone\lab}{2}
{\sin\thetaone\lab} \nonumber \\
&=&  (^{0.2}_{0.3})\, \frac{\qsq+\mxsq}{2}\,f\, (1-\cos\thetaone\lab) \nonumber \\
&=& (^{0.2}_{0.3})\,\frac{\qsq+\mxsq}{2} \, (1-\cos\thetaone\cms)
\eea

\noindent
i.e.

\be
k^2_{\gamma<1} \gtrsim (^{0.2}_{0.3}) k^2_{\gamma=1}
\ee

\vskip 3mm

\noindent
Thus, considering this result along with the dijet virtuality limits
shown in Table\,\ref{table:phillipsvirtuality}, we see that for all
jet production,~$0<\gamma\leq 1$, for a large region of parameter
space, resolved jet production by diagrams such as those shown
in~Figs.\,\ref{fig:dijet} and~\ref{fig:3jets} does not contribute to
the large pseudo-rapidity gap sample.

\subsection*{A.3~~Multi-Jet Production -- Special Case}
\label{sect:lowerquark}

We now consider a scenario which is not covered by either of the cases
described above. This is where the lower parton coupled to the pomeron
emits a gluon, or there is some other QCD radiation, after it interacts
with the pomeron.

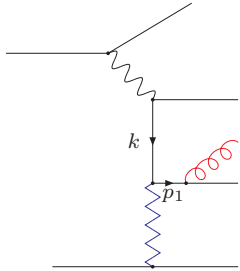
\begin{figure}[htp]
\begin{center} \begin{picture}(180,120)(-20,0)
\SetScale{0.7}
\Line(5,128)(60,128)
\Line(60,128)(105,155)
\Line(84,103)(130,103)
\Text(52,57)[]{\scriptsize $k$}
\ArrowLine(84,58)(102,58)
\Line(102,58)(130,58)
\Text(67,36)[]{\scriptsize $p_{1}$}
\SetColor{Red}
\Gluon(102,58)(130,80){4}{3.5}
\SetColor{Black}
\Vertex(102,58){1}
\ArrowLine(84,103)(84,58)
\Line(30,13)(130,13)
\Photon(60,128)(84,103){4}{3.5}
\SetColor{Blue}
\ZigZag(84,58)(84,13){-4}{4.5}
\SetColor{Black}
\Vertex(60,128){1}
\Vertex(84,103){1}
\Vertex(84,58){1}
\Vertex(84,13){1}
\end{picture}
\caption[Gluon bremsstrahlung in pomeron exchange.]{{\it Gluon
bremsstrahlung in pomeron exchange.}\label{fig:bottomgluon}}
\end{center}
\end{figure}

We consider the case of dijet production, noting that the discussion for
multi-jet production proceeds in an analogous way,  with the
modification that the lower quark coupled to the pomeron emits final
state radiation by single gluon emission. This process is shown in
Fig.\,\ref{fig:bottomgluon}.
 
\noindent
In a resolved-coupling picture, the quark labelled $p_{1}$ should be
close to mass shell. In this case, the quark momentum at the lower
vertex, $p_1$, is shared between the two final-state partons, which,
following
our argument in Appendix\,A.2, means that the jets from the
hadronization of these partons will be more spread than the jet from
one final-state parton carrying all the momentum. The jets from the
final-state partons are all constrained by the pseudo-rapidity cuts to
be separated from the pomeron direction by a minimum angle
$\theta_{\rm min}$, and, since they will spread more than a harder jet,
the angle between the hadronizing partons will be wider and hence the
constraint in this case is stronger than that for a single hard parton
at the lower vertex.

It should be clear from the discussion we present here that the case
for single-gluon bremsstrahlung is sufficient to illustrate the case
for emission of higher numbers of partons from the lower line.

\providecommand\singleletter[1]{#1}


\end{document}